\documentclass[11pt,a4paper,reqno]{amsart}

\usepackage{times}
\usepackage{xspace}
\usepackage[inline]{enumitem}
\usepackage[numbers,sort&compress,longnamesfirst]{natbib}
\usepackage[margin=2.50cm]{geometry}
\usepackage[colorlinks=true, linkcolor=blue,citecolor=blue,urlcolor=blue]{hyperref}
\usepackage[capitalize]{cleveref}

\usepackage{amsfonts,comment}
\usepackage{amsmath}
\usepackage{amssymb,amsmath,latexsym}
\numberwithin{equation}{section}
\usepackage{color,graphicx,colortbl}
\usepackage{xcolor}

\usepackage{amsthm}
\usepackage{graphicx}
\usepackage{multirow}
\usepackage[textwidth=20mm,textsize=tiny]{todonotes}
\usepackage{enumerate}
\usepackage{enumitem}

\usepackage{algpseudocode}  
\usepackage{algorithm}

\definecolor{table_yellow}{HTML}{FFFF00}

\def\<{\langle}
\def\>{\rangle}

\def\e{\mathrm e}  

\newcommand{\be}{\begin{equation}}
\newcommand{\ee}{\end{equation}}
\newcommand{\bea}{\begin{eqnarray}}
\newcommand{\eea}{\end{eqnarray}}
\newcommand{\beas}{\begin{eqnarray*}}
\newcommand{\eeas}{\end{eqnarray*}}
\newtheorem{theorem}{Theorem}[section]

\newtheorem{remark}[theorem]{Remark}
\newtheorem{example}[theorem]{Example}
\newtheorem{examples}[theorem]{Examples}
\newtheorem{foo}[theorem]{Remarks}
\newtheorem{As}{Assumption}

\def\\ud sum{\displaystyle\sum}

\numberwithin{equation}{section}
\numberwithin{figure}{section}
\numberwithin{table}{section}

\makeatletter
\let\c@table\c@figure
\makeatother

\def\ud{\mathrm d}

\begin{document}

\title[Comparison of network architectures]{Machine learning for option pricing:\\an empirical investigation of network architectures}

\author[S. Della Corte]{Serena Della Corte}
\author[L. Van Mieghem]{Laurens Van Mieghem}
\author[A. Papapantoleon]{Antonis Papapantoleon}
\author[J. Papazoglou-Hennig]{Jonas Papazoglou-Hennig}

\address{Delft Institute of Applied Mathematics, TU Delft, 2628 Delft, The Netherlands.}
\email{s.dellacorte@tudelft.nl}

\address{Delft Institute of Applied Mathematics, TU Delft, 2628 Delft, The Netherlands}
\email{laurens@vanmieghem.me}

\address{Delft Institute of Applied Mathematics, TU Delft, 2628 Delft, The Netherlands \& Institute of Applied and Computational Mathematics, FORTH, 70013 Heraklion, Greece}
\email{a.papapantoleon@tudelft.nl}

\address{Department of Mathematics, TU Munich, 85748 Garching, Germany}
\email{jonas.papazoglou-hennig@tum.de}

\thanks{AP gratefully acknowledges the financial support from the Hellenic Foundation for Research and Innovation Grant No. HFRI-FM17-2152. JPH gratefully acknowledges the hospitality at the Financial Engineering \& Mathematical Optimization Lab of the NTUA where this project was initiated.
}



\date{}

\begin{abstract}
We consider the supervised learning problem of learning the price of an option or the implied volatility given appropriate input data (model parameters) and corresponding output data (option prices or implied volatilities).
The majority of articles in this literature considers a (plain) feedforward neural network architecture in order to connect the neurons used for learning the function mapping inputs to outputs.
In this article, motivated by methods in image classification and recent advances in machine learning methods for PDEs, we investigate empirically whether and how the choice of network architecture affects the accuracy and training time of a machine learning algorithm.
We find that highway-type network architectures achieve the best performance, when considering the mean squared error and the training time as criteria, within the considered parameter budgets for the Black–Scholes and Heston option pricing problems. 
Considering the transformed implied volatility problem, a simplified DGM variant achieves the lowest error among the tested architectures. 
We also carry out a capacity-normalised comparison for completeness, where all architectures are evaluated with an equal number of parameters.
Finally, for the implied volatility problem, we additionally include experiments using real market data.
\end{abstract}

\maketitle


\section{Introduction}
\label{sec:intro}

Machine learning has taken the field of mathematical finance by a storm, and there are numerous applications of machine learning in finance by now.
Concrete applications include, for example, the computation of option prices and implied volatilities as well as the calibration of financial models (\citet{pricing_options_implied_vol,MR4188878}), approaches to hedging (\citet{MR3977742}), portfolio selection and optimization (\citet{ZZR}), risk management (\citet{MR4397928}), optimal stopping problems (\citet{MR3960928}), model-free and robust finance (\citet{MR4239795}), stochastic games and optimal control problems (\citet{MR4218407}), as well as the solution of high-dimensional PDEs (\citet{MR3874585}).
A comprehensive overview of machine learning applications in mathematical finance appears in the recent volume of \citet{capponi_lehalle_2023}, while an exhaustive overview focusing on pricing and hedging appears in \citet{RW_JCF}.

We are interested in the computation of option prices and implied volatilities using machine learning methods, and thus, implicitly, in model calibration as well.
More specifically, we consider the supervised learning problem of learning the price of an option or the implied volatility given appropriate input data (model parameters) and corresponding output data (option prices or implied volatilities).
The majority of articles in this literature, see \textit{e.g.} \citet{MR4188878,risks8040101,pricing_options_implied_vol}, consider a (plain) feedforward neural network architecture in order to connect the neurons used for learning the function mapping inputs to outputs.
In this article, motivated by methods in image classification, see \textit{e.g.} \citet{he2015delving,srivastava2015highway,residual_networks}, and recent advances in machine learning methods for PDEs, see \textit{e.g.} \citet{MR3874585}, we investigate empirically whether and how the choice of network architecture affects the speed of a machine learning algorithm, and we are interested in the ``optimal'' network architecture for these problems.

More specifically, next to the classical feedforward neural network or multilayer perceptron (MLP) architecture, we consider residual neural networks, highway networks and generalized highway networks. 
These network architectures have been successfully applied in image classification problems, see \textit{e.g.} \citet{residual_networks}.
Moreover, motivated by the recent work of \citet{MR3874585} on the deep Galerkin method (DGM) for solving high-dimensional PDEs, we also consider the DGM network architecture, and, in addition, we construct and test two variants of this architecture.

The empirical results show that the more advanced network architectures consistently outperform the MLP architecture, and should thus be preferred for the computation of option prices and implied volatilities.
More specifically, we find that for option pricing problems, where we focus on the Black--Scholes and the Heston model, highway-type network architectures outperform all other variants, when considering the mean squared error and the computational time as criteria.
Moreover, we find that for the computation of the implied volatility, after a necessary transformation, a variant of the DGM architecture outperforms all other variants, when considering again the mean squared error and the computational time as criteria.
See also \citet{van_Mieghem_2021}. 

The remainder of this paper is organized as follows: in Section \ref{sec:models} we briefly introduce the models used in this article, and in Section \ref{sec:network_fundamentals} we outline the basics of neural networks.
In Section \ref{sec:network_variations} we describe the different network architectures that will be used throughout, namely residual networks, highway and generalized highway networks, as well as the DGM network and its two variants.
In Section \ref{sec:network_analysis} we revisit the main problems and describe the method used for generating sample data.
In Section \ref{sec:empirical_analysis} we present the results of the empirical analysis performed, where the different network architectures are compared, with main criteria being accuracy and computational time. 
Moreover, we perform a capacity-normalised comparison in which all architectures are evaluated with an equal number of parameters, and also apply our results to real market data.
Finally, Section \ref{sec:conclusion} contains a synopsis and some concluding remarks from the empirical analysis.


\section{Models}
\label{sec:models}

In this brief section, we are going to review the models we will use for option pricing and recall the definition of implied volatility, in order to fix the notation for the remainder of this work.

Let $(\Omega, \mathcal F, \mathbb F, \mathbb Q)$ denote a complete stochastic basis, where $\mathbb F=(\mathcal F_t)_{t\in[0,T]}$ is the filtration, $T>0$ is a fixed and finite time horizon, while $\mathbb Q$ denotes an equivalent martingale measure for an asset $S$ with price process $S=(S_t)_{t\in[0,T]}$, adapted to the filtration $\mathbb F$.  

We first consider the Samuelson / \citet{black_scholes_pde} model for the evolution of the asset price $S$, \textit{\textit{i.e.}}
\begin{equation}
	\label{eq:dynamics_s_real_world}
	\ud S_t = r S_t \ud t + \sigma S_t \ud W_t, \quad S_0>0,
\end{equation}
where $r$ denotes the risk-free interest rate, $\sigma$ the volatility and $W$ denotes a $\mathbb Q$-Brownian motion.
The price of a European call option with payoff $C=(S_T-K)^+$ in this model is provided by the celebrated Black--Scholes formula, \textit{\textit{i.e.}}
\begin{align}
	\label{eq:call_option_analytical_solution}
	  	\begin{split}
			\pi_t(C;K,T,\sigma) &= S_t \Phi(d_1) - \e^{-r(T-t)} K \Phi(d_2),\\
			d_1 & = \frac{\log \big(\frac{S_t}{K}\big) + (r + \frac{1}{2}\sigma^2)(T - t)}{\sigma\sqrt{T - t}}
				  = d_2 + \sigma \sqrt{T - t},
		\end{split}		  
\end{align}
where $\Phi(\cdot)$ denotes the cumulative distribution function (cdf) of the standard normal distribution. 

\begin{figure}[h!]
	\centering
	\includegraphics[width=0.8\linewidth]{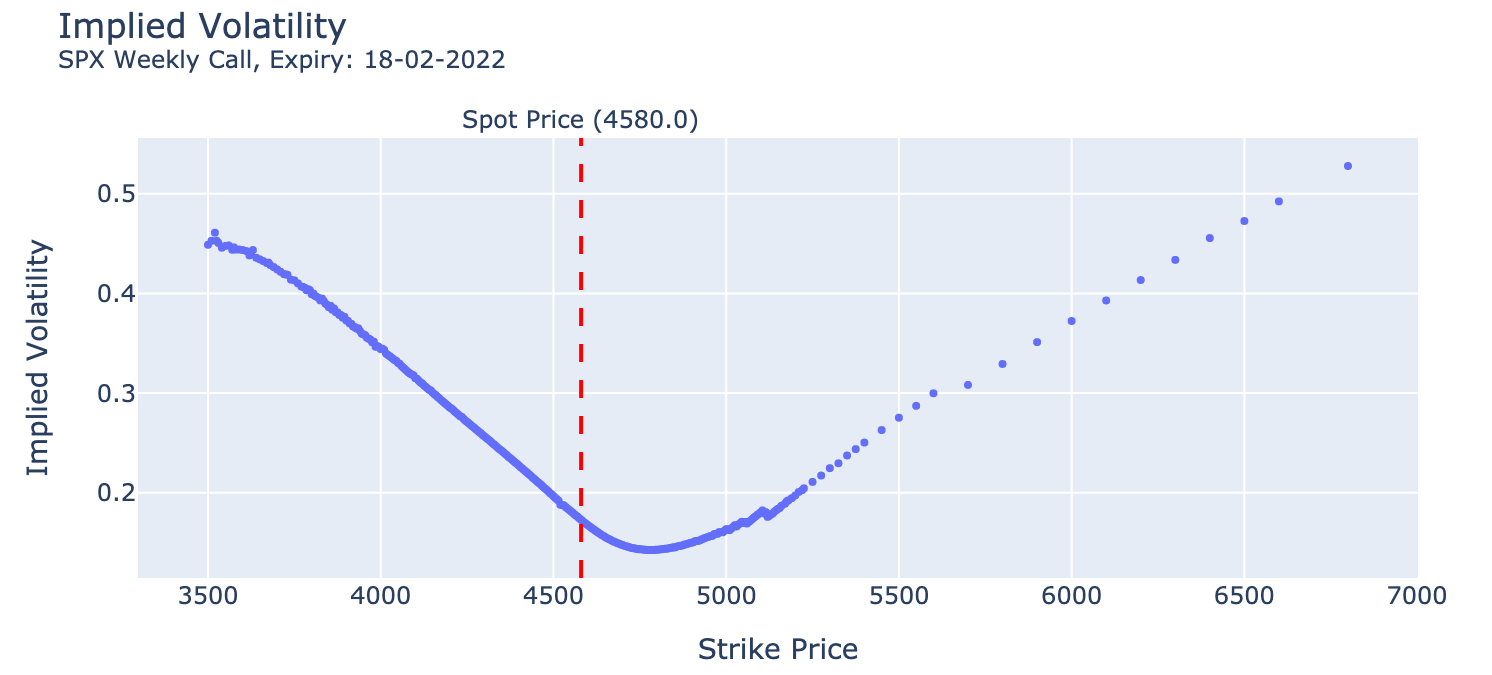}
	\caption{The market implied volatility for CBOE's SPX options.}
	\label{fig:implied_volatility_skew}
\end{figure}

Assume that we can observe in a financial market the traded prices $\pi_t^{\mathrm{M}}(C;K,T)$ for a call option $C$, for various strikes $K$ and maturities $T$.
The implied volatility $\sigma_{\mathrm{iv}}(K,T)$ is the volatility that should be inserted into the Black--Scholes equation \eqref{eq:call_option_analytical_solution} such that the model price $\pi_t(C;K,T,\sigma)$ matches the traded market price $\pi_t^{\mathrm{M}}(C;K,T)$, \textit{\textit{i.e.}}
\begin{equation}
	\label{eq:ivol}
		\pi_t(C;K,T,\sigma_{\mathrm{iv}}(K,T)) = \pi_t^{\mathrm{M}}(C;K,T),
\end{equation}
see \textit{e.g.} \citet[Ch. 1]{Gatheral_2006}.
The inverse problem in \eqref{eq:ivol} does not admit a closed form solution and needs to be solved numerically, using methods such as the Newton--Raphson or Brent; see \textit{e.g.} \citet{pricing_options_implied_vol} for more details. 
The Black--Scholes model assumes that the volatility is constant; however the implied volatility computed from real market option prices exhibits a so-called smile or skew shape; see \textit{e.g.} Figure \ref{fig:implied_volatility_skew} for a visualization.

A popular model that captures the behavior of market data is the Heston stochastic volatility model, \textit{cf.} \citet{Heston93}.
The dynamics of the asset price in this model are provided by the following SDEs
\begin{align}
\label{eq:Heston_model}
	\begin{cases}
		\ud S_t = rS_t \ud t + \sqrt{v_t}S_t \ud W_t,  & S_0 > 0 \\
		\ud v_t = \kappa (\theta - v_t) \ud t + \eta \sqrt{v_t} \ud \widetilde W_t, & v_0 > 0,
	\end{cases}
\end{align}
where $\kappa, \theta, \eta$ denote positive constants and $[W,\widetilde W] = \rho$ denotes the correlation between the two $\mathbb Q$-Brownian motions. 
The tuple of the log-asset price process and the volatility process $(\log S, v)$ is an affine process, see \textit{e.g.} \citet[Ch. 10]{Filipovic09}, and the characteristic function of the log-asset price can be computed explicitly.
Therefore, option prices in the Heston model can be computed efficiently using Fourier transform methods, see \textit{e.g.} \citet{fourier_transform_valuation}, or the COS method of \citet{Fang_Oosterlee_2008}.

\section{Fundamentals of neural networks}
\label{sec:network_fundamentals}

In this section, we briefly recall the basic building blocks of feedforward neural networks, fix notation, and introduce the multilayer perceptron (MLP) architecture used as a reference throughout the paper.

A (single) \emph{perceptron} is a function that maps an input vector $\mathbf{x}\in\mathbb{R}^d$ to an output $\mathbf{y}\in\mathbb{R}$ of the form
\begin{equation}
	\label{eq:output_perceptron_function}
	\mathbf{y} = H\left( \sum_{i=1}^{d} w_i x_i  + b \right)
	= H\left(\mathbf{w}\cdot \mathbf{x} + b\right),
\end{equation}
where $\mathbf{w}\in\mathbb{R}^d$ is a weight vector, $b\in\mathbb{R}$ a bias term, and $H$ is a nonlinear activation function. 
Here $\cdot$ denotes the standard inner product.

The activation function $H$ is typically nonlinear (to allow the network to represent nonlinear maps) and differentiable (to enable gradient-based training). 
Common examples include:
\begin{itemize}
	\item the sigmoid function, $H_{\rm sig}(x) = \frac{1}{1 + \e^{-x}}$,
	\item the hyperbolic tangent, $H_{\rm tanh}(x) = \frac{\e^x - \e^{-x}}{\e^x + \e^{-x}}$,
	\item the Rectified Linear Unit (ReLU), $H_{\rm ReLU}(x) = \max\{0, x\}$,
	\item the Gaussian Error Linear Unit (GELU), $H_{\rm GELU}(x) = x \,\Phi (x)$, with $\Phi$
	      the standard normal cdf,
	\item the softmax function for classification,
	      $H_{\rm sm}(\mathbf{x})_i = \frac{\e^{x_i}}{\sum _{j=1}^K \e^{x_j}}$.
\end{itemize}

A \emph{layer} is a collection of perceptrons that share the same input. 
Let a layer have $n$ units, and collect their outputs into a vector $\mathbf{y}\in\mathbb{R}^n$. 
Using a weight matrix $W\in\mathbb{R}^{n\times d}$ and bias vector $\mathbf{b}\in\mathbb{R}^n$, we can write
\begin{equation}
\label{eq:MLP}
	y_i = H_i \left( \sum _{j=1}^d W_{ij} x_j + b_i \right),
\end{equation}
where $H_i$ is the activation of unit $i$. 
In most applications, the same activation function is used throughout a layer, hence $H_i \equiv H$ for all $i$.

\begin{figure}[h!]
	\centering
	\includegraphics[scale=0.5]{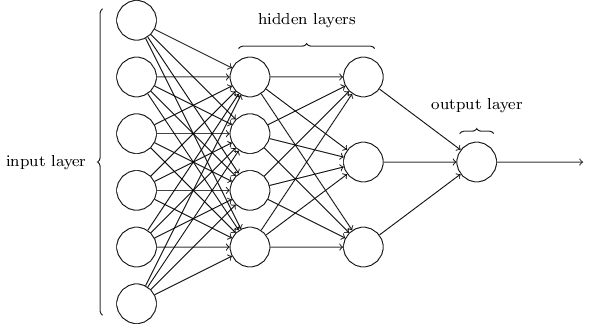}
	\caption{A visual representation of a multilayer perceptron (MLP).}
	\label{fig:multilayer_perceptron}
\end{figure}

The simplest example of a feedforward neural network is the \emph{multilayer perceptron} (MLP), where layers are stacked and fully connected: the output of one layer serves as the input to the next; see Figure~\ref{fig:multilayer_perceptron}.

Consider a target output $\mathbf{y}$ and a neural network $H(\mathbf{x};\theta)$ with parameters $\theta$ (all weights and biases).
We measure the discrepancy by a loss function $J$, \textit{e.g.}\ the mean squared error,
\begin{equation}
\label{eq:loss_function}
	J(\theta) := \| H(\mathbf{x}; \theta) - \mathbf{y}\|^{\color{red}2}.
\end{equation}
The parameters are learned by (stochastic) gradient descent,
\begin{equation}
\label{eq:updating_weights}
	\theta_{k+1} := \theta_k - \alpha \nabla _{\theta} J(\theta_k),
\end{equation}
with learning rate $\alpha>0$. In practice, the training data are partitioned into mini-batches; the update~\eqref{eq:updating_weights} is applied on each batch, and one full pass over all batches is referred to as an \emph{epoch}. 
We initialize all networks using either the Glorot \cite{glorot_initializer} or the He \cite{he2015delving} schemes.


\section{Network variations}
\label{sec:network_variations}

Let us now describe the network architectures used in our experiments. 
All of them are based on the fully-connected MLP structure from Section~\ref{sec:network_fundamentals}, and differ only on how each layer transforms its input, and how information is propagated across layers.
Throughout this section, we follow the notation of \citet{srivastava2015highway}.

\begin{figure}[h!]
	\centering
	\includegraphics[width=0.55\linewidth]{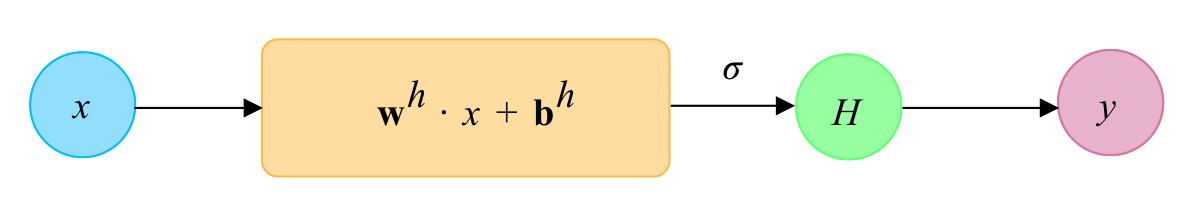}
	\caption{Schematic representation of a single MLP layer, \textit{cf.}~\eqref{eq:MLP_layer_operation}.}
	\label{fig:MLP_layer_schematic}
\end{figure}

A standard MLP layer applies an affine map followed by a nonlinearity,
\begin{equation}
\label{eq:MLP_layer_operation}
	\mathbf{y} = H(W_H \mathbf{x} + \mathbf{b}_H) =: H(\mathbf{x}, W_H),
\end{equation}
where $W_H$ and $\mathbf{b}_H$ denote the layer-specific weights and biases. 
A schematic representation is shown in Figure~\ref{fig:MLP_layer_schematic}; the symbol $\sigma$ is used to indicate a generic nonlinear activation function.

\subsection{Residual network}
\label{subsec:residual_network}

\begin{figure}[h!]
	\centering
	\includegraphics[width=0.8\linewidth]{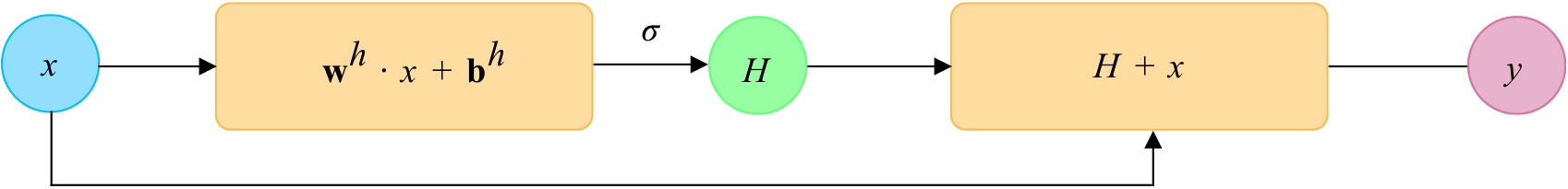}
	\caption{Schematic representation of a single residual layer, \textit{cf.}~\eqref{eq:residual_layer_operation}.}
	\label{fig:residual_layer_schematic}
\end{figure}

Residual networks (ResNets; \citet{residual_networks}) augment the MLP layer with an identity skip connection, so that each layer computes
\begin{equation}
\label{eq:residual_layer_operation}
	\mathbf{y} = H(\mathbf{x}, W_H) + \mathbf{x}.
\end{equation}
A schematic view appears in Figure~\ref{fig:residual_layer_schematic}. 
The additional skip connection allows information (and gradients) to propagate directly across layers and has been shown to improve trainability in deep architectures.

\subsection{Highway network}
\label{subsec:highway_networks}

\begin{figure}[h!]
	\centering
	\includegraphics[width=0.8\linewidth]{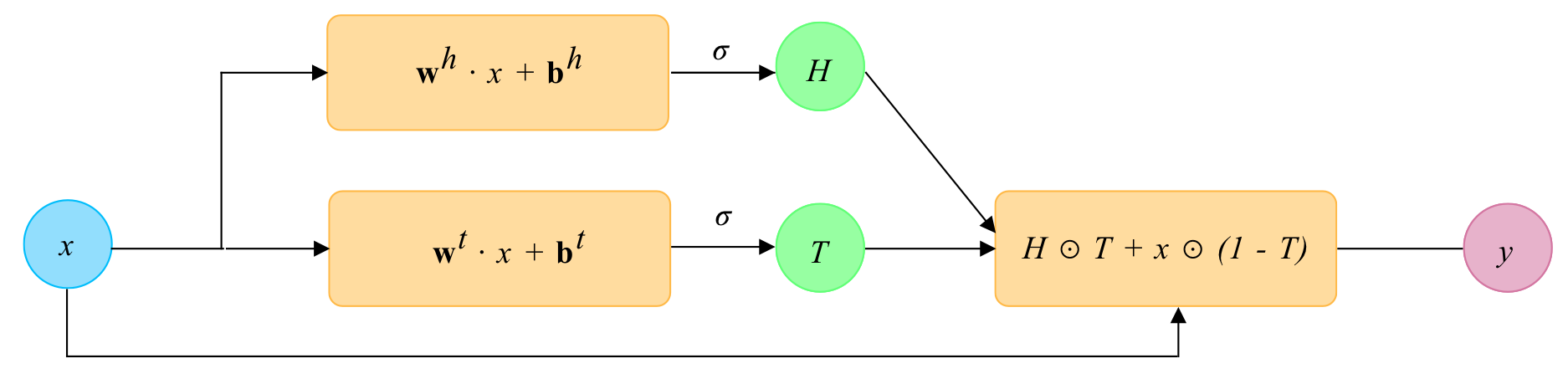}
	\caption{Schematic representation of a single highway layer, \textit{cf.} \eqref{eq:simplified_highway_layer}.}
	\label{fig:highway_layer_schematic}
\end{figure}

Highway networks are a generalization of the residual network architecture from the previous subsection, by introducing an additional non-linearity $T(\mathbf{x}, W_T)$, called the transform gate. 
This non-linearity will determine the amount of information flowing from the non-linearity $H$ inside the layer and the amount of information `carried' over from the input vector $\mathbf{x}$. 
The outcome is thus the following transformation
\begin{equation}
\label{eq:simplified_highway_layer}
	\mathbf{y} = H(\mathbf{x}, W_H) \odot T(\mathbf{x}, W_T) + \mathbf{x} \odot \big[1 - T(\mathbf{x}, W_T)\big],
\end{equation}
where $\odot$ denotes the Hadamard product between vectors or matrices.
Hence, $T$ creates a convex combination between the input $\mathbf{x}$ and the transformed input $H(\mathbf{x}, W_H)$ before the output. 
A visual representation of this layer is given in Figure \ref{fig:highway_layer_schematic}.

In the approximation of complex functions using neural networks, the depth of the network is proven to be a significant factor in its success, see \textit{e.g.} \citet{Telgarsky2016BenefitsOD}. 
However, training deeper networks presents additional bottlenecks, such as the vanishing gradients problem. 
The intuition behind highway networks is to allow for unimpeded information flow across layers; see \citet{srivastava2015highway}.
Indeed, from equation \eqref{eq:simplified_highway_layer} we can see that for the boundary values of $T$ we have 
\begin{equation}
	\mathbf{y} = 
	\begin{cases} 
      \mathbf{x}, & T(\mathbf{x}, W_T) = \mathbf{0} \\
	  H(\mathbf{x}, W_H), & T(\mathbf{x}, W_T) = \mathbf{1},
   \end{cases}
\end{equation}
therefore
\begin{equation}
	\frac{\ud\mathbf{y}}{\ud\mathbf{x}} = 
	\begin{cases} 
      I, & T(\mathbf{x}, W_T) = \mathbf{0} \\
	  H'(\mathbf{x}, W_H), & T(\mathbf{x}, W_T) = \mathbf{1},
   \end{cases}
\end{equation}
where $I$ denotes the identity matrix. 
Thus, the transform gate $T$ allows the highway layer to vary its behaviour between a standard MLP layer and an identity mapping, leaving $\mathbf{x}$ unaffected. 


\subsection{Generalized highway network}
\label{subsec:generalized_highway_network}

\begin{figure}[h!]
	\centering
	\includegraphics[width=0.7\linewidth]{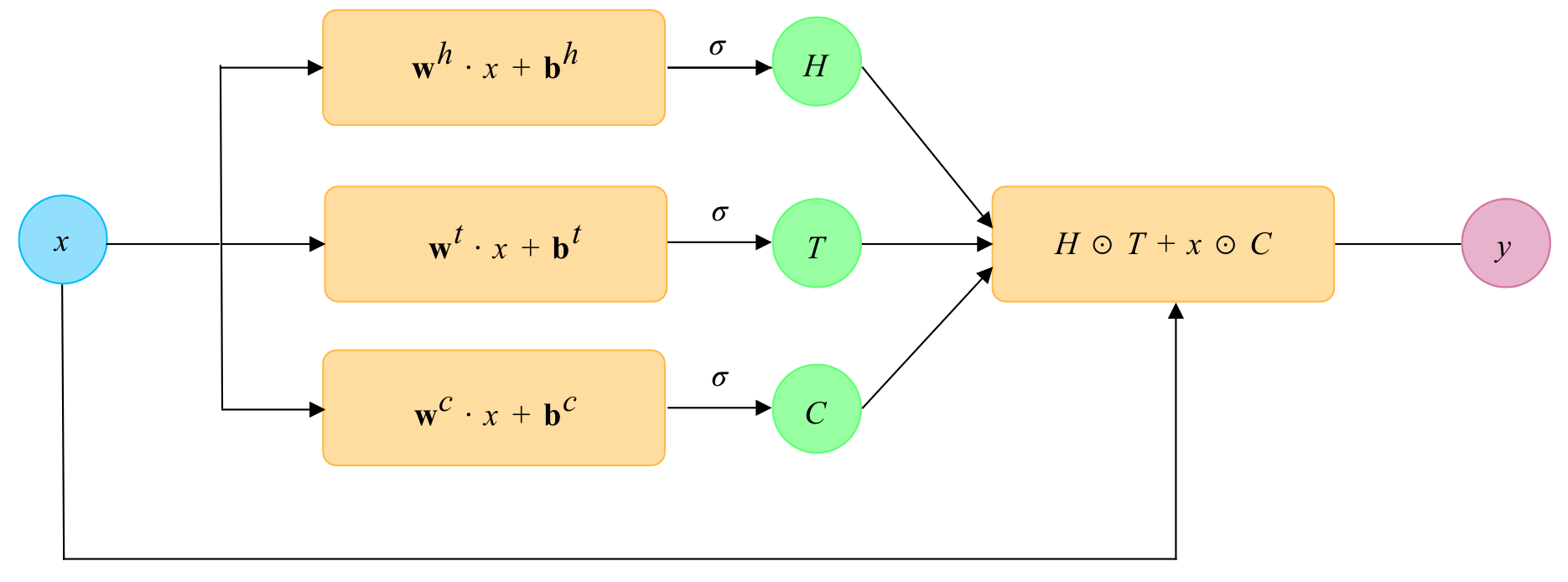}
	\caption{Schematic representation of a single generalized highway layer, \textit{cf.} \eqref{eq:general_highway_layer}.}
	\label{fig:generalized_highway_layer_schematic}
\end{figure}

One can further generalize highway networks by removing the convexity requirement in the outcome. 
Let us define an additional non-linear transformation alongside $T(\mathbf{x}, W_T)$, which we denote by $C(\mathbf{x}, W_C)$, such that the outcome takes the form
\begin{equation}
\label{eq:general_highway_layer}
	\mathbf{y} = H(\mathbf{x}, W_H) \odot T(\mathbf{x}, W_T) + \mathbf{x} \odot C(\mathbf{x}, W_C).
\end{equation} 
We call $T$ and $C$ the transform and carry gate, respectively, as they regulate the amount of information passed into the next layer from the transformed and original input. 
This gating mechanism allows information to flow along the layers of the network without degradation.
Highway networks are obviously a special case of generalized highway networks, once we set $C=1-T$. 
A visual representation of this layer is given in Figure \ref{fig:generalized_highway_layer_schematic}.

\subsection{DGM network}
\label{subsec:DGM_network}



The DGM network architecture was proposed in the context of the Deep Galerkin Method for the solution of PDEs; see \citet{MR3874585} and also \citet{alaradi2018solving}. 
This layer structure is related to generalized highway networks but incorporates additional nonlinear blocks and an explicit dependence on the original input $\mathbf{x}$ at each layer.


\subsubsection{DGM layer}
\label{subsec:dgm_layer}

\begin{figure}[h!]
	\centering
	\includegraphics[width=0.8\linewidth]{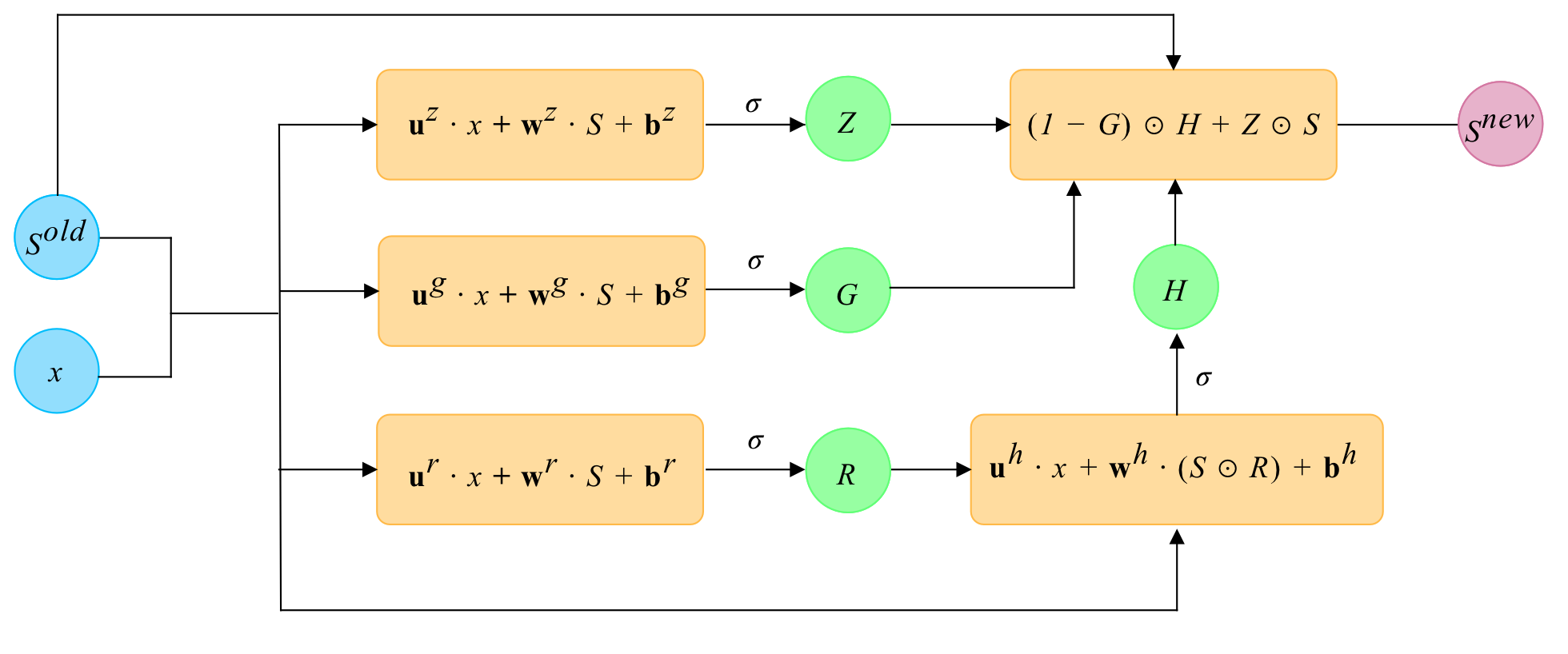}
	\caption{Schematic representation of a single DGM layer. Adopted from \cite{alaradi2018solving}.}
	\label{fig:dgm_layer_schematic}
\end{figure}

A DGM layer takes as input the feature vector $\mathbf{x}$ and the output $S^{\mathrm{old}}$ of the previous layer (or an initial affine transform of $\mathbf{x}$). 
Using these, it computes four intermediate vectors $Z,G,R,H$ and produces the new state $S^{\mathrm{new}}$ via
\begin{equation}
\label{eq:final_dgm_operation}
	S^{\mathrm{new}} = (1 - G) \odot H + Z \odot S,
\end{equation}
where $S$ denotes the incoming state. Setting $T = 1-G$ and $C = Z$ shows that this is formally analogous to a generalized highway layer,
\[
	S^{\mathrm{new}} = T \odot H + C \odot S,
\]
but with a more elaborate internal computation of $H$ and recurrent use of the original input $\mathbf{x}$; see Figure~\ref{fig:dgm_layer_schematic}.

Compared to a highway layer, a DGM layer is more complex in two respects:
\begin{itemize}
	\item[(i)] The vector $H$ is obtained by composing multiple nonlinear transformations, which can be interpreted as a small ``subnetwork'' within the layer.
	\item[(ii)] The original feature vector $\mathbf{x}$ enters each nonlinear block, creating a recurrent-like structure reminiscent of gated recurrent units; see, \textit{e.g.},
	          \citet{Sherstinsky_2020}.
\end{itemize}
These two aspects will be further explored in the next subsections, by introducing and testing new architectures that stress these points.
As a consequence, a DGM layer contains substantially more parameters than an MLP or a highway layer and incurs higher computational cost, but offers increased flexibility for approximating functions with sharp local features.

\subsubsection{Overall DGM architecture}
\label{subsec:dgm_overall}

\begin{figure}[h!]
	\centering
	\includegraphics[width=0.8\linewidth]{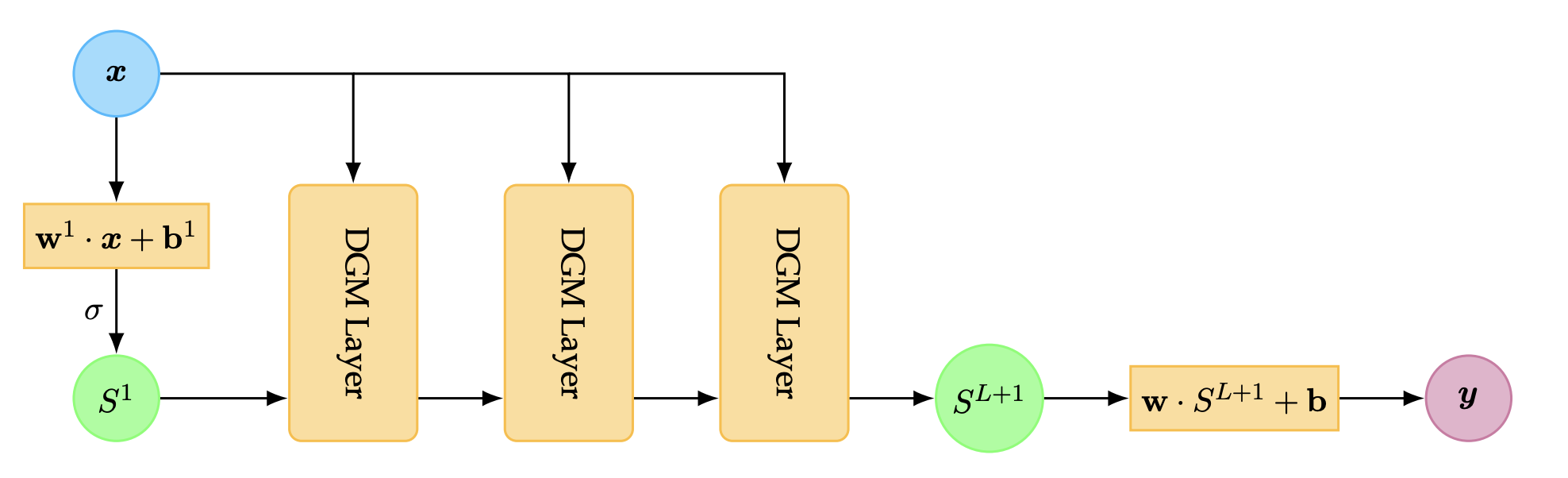}
	\caption{High-level overview of the DGM network. Each block labelled ``DGM layer''
	         refers to the operations in Figure~\ref{fig:dgm_layer_schematic}. Adopted from
	         \cite{alaradi2018solving}.}
	\label{fig:dgm_model_schematic}
\end{figure}

Since each DGM layer requires both $\mathbf{x}$ and a hidden state $S$ as input, we first compute an initial state $S^1$ via a dense layer,
\begin{equation}
	S^1 = \mathbf{w}^1 \cdot \mathbf{x} + \mathbf{b}^1,
\end{equation}
which sets the dimension of the subsequent DGM layers. 
An output layer then maps the final state to the desired prediction. 
The overall architecture is summarized in Figure~\ref{fig:dgm_model_schematic}.

\subsection{Deep DGM network}
\label{subsec:deep_dgm_layer}

\begin{figure}[h!]
	\centering
	\includegraphics[width=0.8\linewidth]{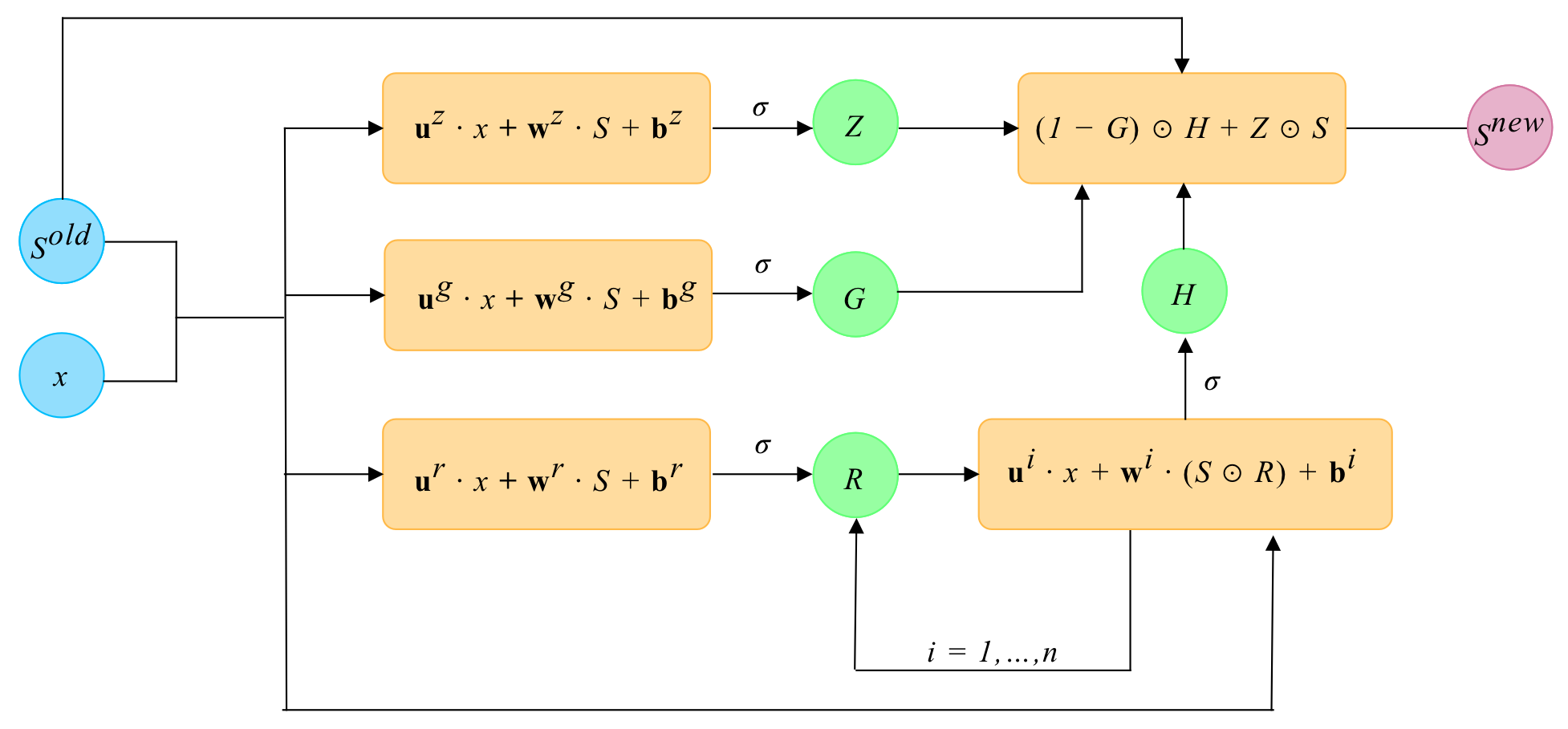}
	\caption{Schematic representation of a deep DGM layer with $n$ internal nonlinear transformations.}
	\label{fig:deep_dgm_layer}
\end{figure}

The \emph{deep DGM} variant increases the depth of the internal subnetwork used to compute $H$, by stacking $n$ nonlinear transformations; see Figure~\ref{fig:deep_dgm_layer}. 
This adds additional weight matrices and biases inside each DGM layer and allows us to study the effect of deeper intra-layer nonlinearities on approximation quality.

\subsection{No-Recurrence DGM network}
\label{subsec:no_recurrence_dgm_network}

\begin{figure}[h!]
	\centering
	\includegraphics[width=0.8\linewidth]{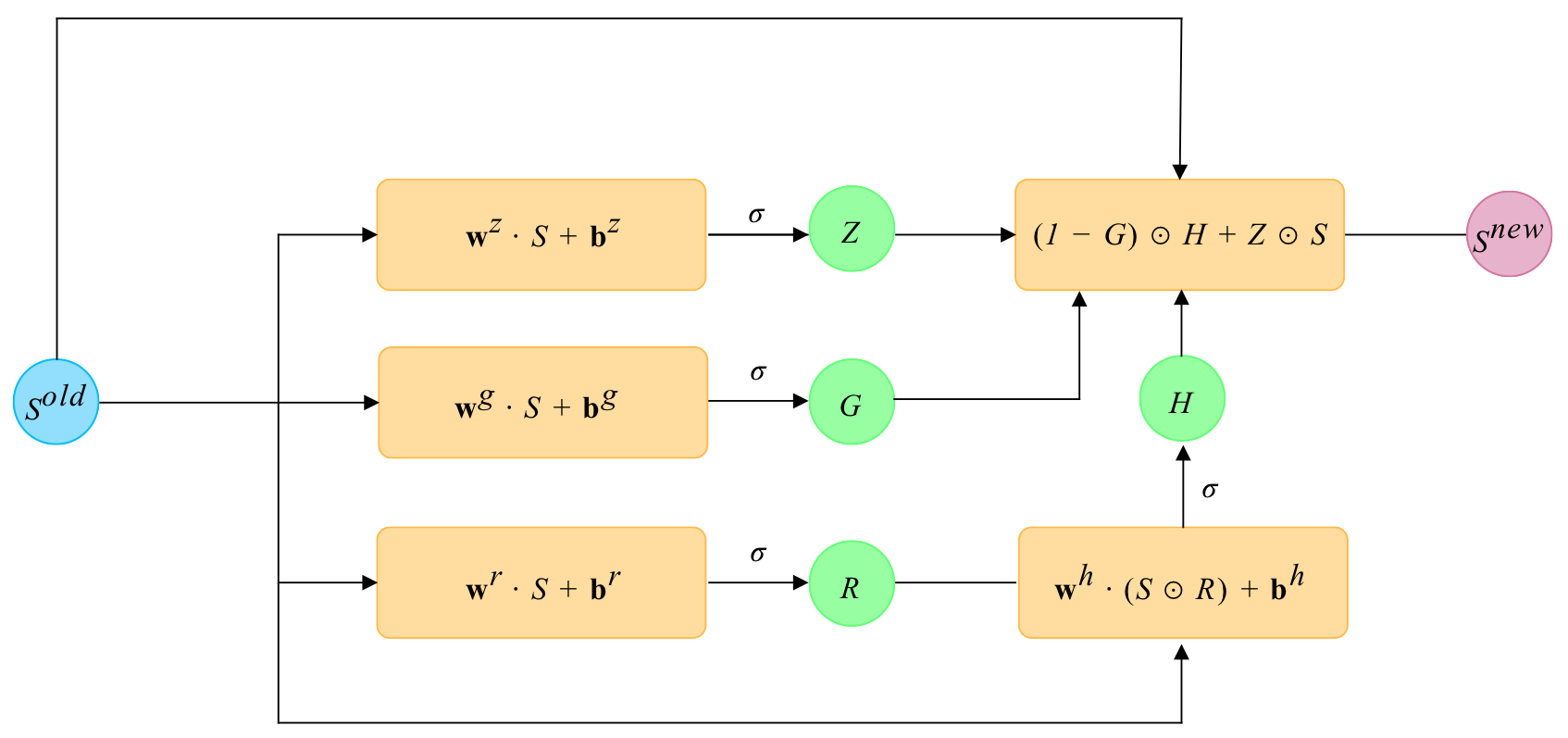}
	\caption{Schematic representation of a single no-recurrence DGM layer.}
	\label{fig:dgm_no_recurrence}
\end{figure}

Finally, we consider a simplified \emph{no-recurrence} DGM network in which the dependence on $\mathbf{x}$ is removed from all layers except the first. 
The resulting layer structure, shown in Figure~\ref{fig:dgm_no_recurrence}, closely resembles a highway layer but retains the more complex computation of $H$:
\begin{equation}
	H = \sigma \Big( \mathbf{w}^h \cdot \big[S \cdot \sigma\left(\mathbf{w}^r \cdot S + \mathbf{b}^r\right) \big]
		+ \mathbf{b}^h \Big).
\end{equation}
This reduces the parameter count and computational cost per layer, while preserving some of the additional flexibility of the original DGM construction.

\section{Problems revisited and data sampling}
\label{sec:network_analysis}

In this section, we take another look at the problems considered in this work and present the form that will be used in the empirical analysis.
We also discuss the methods for generating data and the parameter sets used for training the neural networks.


\subsection{Black--Scholes and Heston models}

The first problem we consider is the pricing of European call options in the Black--Scholes model. 
Let us define the following variables, from which we will sample later:
\begin{itemize}
	\item $m = \frac{S}{K}$: the moneyness value;
	\item $\sigma$: the volatility of the underlying asset;
	\item $r$: the risk-free rate;
	\item $\tau = T - t$: a time transformation to obtain the time to maturity, in years.
\end{itemize}
Using these variables, we can calculate the price of a European call option using a transformed version of \eqref{eq:call_option_analytical_solution}:
\begin{align}
\label{eq:analytical_solution_black_scholes_transformed}
  	\begin{split}
		\pi_t(C;m,\tau, \sigma) &= \frac{\pi_t(C; K, T, \sigma)}{K} = m\Phi(d_1) - \e^{-r\tau}\Phi(d_2), \\
		d_1 &= \frac{\log (m) + (r + \frac{1}{2}\sigma^2)\tau}{\sigma\sqrt{\tau}} = d_2 + \sigma \tau.
	\end{split}
\end{align}

\noindent When we train neural networks to compute call option prices in the Black--Scholes model, we will use the solution presented in \eqref{eq:analytical_solution_black_scholes_transformed}. 

In addition to the Black--Scholes pricing problem, we also train neural networks to compute the price of European call options in the Heston stochastic volatility model \eqref{eq:Heston_model}.
The prices of call options in the Heston model are computed using the COS method and utilizing the affine structure of the characteristic function.
We use this problem in order to investigate whether the various networks show consistent performance across different option pricing problems or not. 


\subsection{Implied volatility}
\label{subsec:implied_volatility_problem_description}

The next problem we consider is the training and evaluation of neural networks for the computation of the implied volatility, thus continuing the analysis of \citet{pricing_options_implied_vol}, in which only MLPs are trained and evaluated.
In addition to the standard dataset for the computation of the implied volatility, in which we use the parameters directly from the inverted Black--Scholes pricing problem (Table \ref{table:implied_volatility_parameters}), we also apply a transformation on the dataset from the scaled call price $(\pi/K)$ to the scaled time value of the option, already introduced in \citet{pricing_options_implied_vol}. 
The reasoning behind this transformation is the following: 
We train the networks to find the solution to equation \eqref{eq:ivol}. 
However, we know that the option's vega, \textit{i.e.} its derivative with respect to the volatility, can become arbitrarily small for deep in-the-money (ITM) or deep out-of-the-money (OTM) options. 
In the context of neural networks, where we take the derivatives to compute the gradient of the network, the instability of vega may lead to large gradients, possibly causing significant estimation errors. 

Let us briefly elaborate further on this issue. 
We denote $\pi =: f(\sigma)$ for convenience, where $f$ is the analytical function representing the price of European call option in the Black--Scholes model, depending on $\sigma$. 
The implied volatility problem concentrates on approximating the inverse of this problem, \textit{i.e.} $\sigma = f^{-1}(\pi)$. 
During the backpropagation phase, we must compute the gradient of $f^{-1}(\pi)$ with respect to $\pi$. 
That is, $\frac{d\sigma}{d\pi} = \frac{1}{d\pi/d\sigma}$, which is the reciprocal of the vega. 
Hence, for arbitrarily small values of vega, this gradient will explode, leading to convergence problems. 

In order to address this issue, we will follow the gradient-squash approach of \citet{pricing_options_implied_vol}.
An option value can be divided into its \textit{intrinsic value} (\textit{i.e.} the no-arbitrage bound) and its \textit{time value}. 
In order to obtain the time value, we subtract the intrinsic value from the option value, yielding
\begin{equation}
\label{eq:scaled_time_value_transformation}
	\hat{\pi}_t = \pi_t - (S_t - K\e^{-r\tau})^+,
\end{equation}  
where $\hat{\pi}$ denotes the time value. 
Finally, we will also apply a $\log$-transform to further reduce the possible steepness of the gradient, resulting in the dataset shown in Table \ref{table:transformed_implied_volatility_parameters}. 


\subsection{Data sampling}

We treat each problem discussed above as a \textit{supervised learning} problem. 
In other words, we require inputs and outputs which we can feed to the network and calculate the loss function. 
We will follow the configuration outlined in \citet{pricing_options_implied_vol} for all three problems. 
We will generate a total of $1$ million samples for each problem, which we divide into an $80/20$ training and validation set. 
We will generate an additional $100$ thousand samples used for evaluation. 
When sampling from the space of input parameters, we can either define a joint distribution over the entire domain, or sample each variable separately. 
\citet{pricing_options_implied_vol}, as well as \citet{jph-report} and \citet{van_Mieghem_2021}, opt for Latin hypercube sampling (LHS), in which values are sampled from a joint distribution, resulting in a better representation of the parameter space; \textit{cf.} \citet{latin_hypercube_sampling}. 
Thus, in order to remain consistent with previous research, we use LHS in this work to generate the data as well.

Starting from the Black--Scholes pricing problem, we sample points using the parameters shown in Table \ref{table:black_scholes_parameters}, and use the analytical solution in equation \eqref{eq:analytical_solution_black_scholes_transformed} to obtain the output labels.
We use the same sampling parameters as in \cite{pricing_options_implied_vol, jph-report}, allowing us to compare the network performances with the results of both works at a later stage. 

\begin{table}[H]
\centering
\begin{tabular}{@{} l|c|c @{}}
  \hline
  & Parameters & Range \\
  \hline
  \multirow{4}{*}{Input}  & Moneyness: $S_0 / K$ & $[0.4, 1.6]$ \\
	     & Time to Maturity ($\tau = T - t$) & $[0.2, 1.1]$ \\
         & Risk-free rate ($r$) & $[0.02, 0.1]$ \\
         & Volatility ($\sigma$) & $[0.01, 1.0]$ \\
   \hline
  Output   & Scaled call price ($\pi / K$) & $(0.0, 0.9)$ \\
  \hline
\end{tabular}
\caption{Black--Scholes data generating parameters, adopted from \cite{pricing_options_implied_vol}.}
\label{table:black_scholes_parameters}
\end{table}

The Heston pricing problem obviously requires additional parameters, whose range is listed in Table \ref{table:heston_parameters}. 
The range for these parameters is selected again according to \citet{pricing_options_implied_vol}.

\begin{table}[H]
\centering
\begin{tabular}{@{} l|c|c @{}}
  \hline
  & Parameters & Range \\
  \hline
  \multirow{8}{*}{Input}  & Moneyness: $S_0 / K$ & $[0.4, 1.6]$ \\
	     & Time to Maturity ($\tau = T - t$) & $[0.2, 1.1]$ \\
         & Risk-free rate ($r$) & $[0.02, 0.1]$ \\
         & Correlation ($\rho$) & $[-0.95, 0.0]$ \\
         & Reversion speed ($\kappa$) & $[0, 2.0]$ \\
         & Long term variance ($\bar{v}$) & $[0, 0.5]$ \\
         & Volatility of volatility ($\gamma$) & $[0, 0.5]$ \\
         & Initial variance ($v_0$) & $[0.05, 0.5]$ \\
   \hline
  Output & Call price ($\pi$) & $(0.0, 0.67)$ \\
  \hline
\end{tabular}
\caption{Heston data generating parameters, adopted from \cite{pricing_options_implied_vol}.}
\label{table:heston_parameters}
\end{table}

In the `default' implied volatility problem, we reuse the Black--Scholes pricing data and switch the role of the volatility $(\sigma)$ and the scaled call price $(\pi/K)$ as input and output, obtaining the parameters listed in Table \ref{table:implied_volatility_parameters}.

\begin{table}[H]
\centering
\begin{tabular}{@{} l|c|c @{}}
  \hline
  & Parameters & Range \\
  \hline
  \multirow{4}{*}{Input}  & Moneyness: $S_0 / K$ & $[0.4, 1.6]$ \\
	     & Time to Maturity ($\tau = T - t$) & $[0.2, 1.1]$ \\
         & Risk-free rate ($r$) & $[0.02, 0.1]$ \\
         & Scaled call price ($\pi / K$) & $[0, 0.71]$ \\
   \hline
  Output   & Volatility ($\sigma$) & $(0.05, 1.0)$ \\
  \hline
\end{tabular}
\caption{Implied volatility data generating parameters, adopted from \cite{pricing_options_implied_vol}.}
\label{table:implied_volatility_parameters}
\end{table}

Finally, we transform the implied volatility dataset by applying the transformation described in equation \eqref{eq:scaled_time_value_transformation}, along with a $\log$-transform. 
The domain of the transformed implied volatility dataset, where we used the scaled time value, is provided in Table \ref{table:transformed_implied_volatility_parameters}.
\begin{table}[H]
\centering
\begin{tabular}{@{} l|c|c @{}}
  \hline
  & Parameters & Range \\
  \hline
  \multirow{4}{*}{Input}  & Moneyness: $S_0 / K$ & $[0.4, 1.6]$ \\
	     & Time to Maturity ($\tau = T - t$) & $[0.2, 1.1]$ \\
         & Risk-free rate ($r$) & $[0.02, 0.1]$ \\
         & Scaled time value ($\log (\hat{\pi} / K)$) & $[-18.42, -0.95]$ \\
   \hline
  Output   & Volatility ($\sigma$) & $(0.01, 1.0)$ \\
  \hline
\end{tabular}
\caption{Transformed implied volatility data generating parameters, adopted from \cite{pricing_options_implied_vol}.}
\label{table:transformed_implied_volatility_parameters}
\end{table}  


\section{Empirical results}
\label{sec:empirical_analysis}

In this section, we train neural networks using the architectures discussed in Sections \ref{sec:network_fundamentals} and \ref{sec:network_variations}, and apply them to the three different problems outlined in Sections \ref{sec:models} and \ref{sec:network_analysis}. 
We then compare the empirical performance of these networks against each other, with the aim of identifying an optimal network architecture for the solution of these and related problems. 
In addition, we test the method for the computation of implied volatility using real data, and also briefly compare this with existing methods.
The implementation for every network was done using Google's \textit{TensorFlow} library for Python \cite{tensorflow2015-whitepaper} and the empirical analysis was carried out using the Delft High Performance Computing Center \cite{DHPC2022} facilities. 


\subsection{Multilayer perceptrons}
\label{subsec:multilayer_perceptron_analysis}

We start by training a set of MLPs ranging from a very small network to a large network, with similar configurations. 
The goal is to evaluate the performance of this network architecture, as well as to determine whether the size of the network influences the performance over a fixed range of epochs. 
To this end, we define a set of twelve networks with configurations listed in Table \ref{table:mlp_set}. 
We fix the learning rate according to the optimal value found from the study conducted in \citet{pricing_options_implied_vol}. 
The batch size is chosen as large as possible while keeping the calculations on a CPU fast. 
We found that a batch size of $64$ works well for the supervised learning problems discussed in this work. 
Furthermore, each network in this work is trained for $200$ epochs on the entire training dataset. 
We found that, in most cases, increasing the number of training epochs is beneficial. 
However, due to the large amount of network architectures that must be trained, we decided to terminate training after $200$ epochs. 

\begin{table}[h!]
\centering
\begin{tabular}{ c|c } 
\hline
 & Multilayer Perceptron (MLP) \\ 
\hline
Layers & $2, 3$ \\
Nodes per layer & $50, 100, 150, 200, 250, 500$ \\
Activation function & ReLU \\
Loss function & MSE \\
Learning rate & $10^{-5}$ \\
Batch size & 64 \\
\hline
\end{tabular}
\caption{Overview of MLP configurations.}
\label{table:mlp_set}
\end{table} 

We allocate exactly one CPU core from the DHPC \cite{DHPC2022} (Intel Xeon 3.0GHz) cluster to each model. 
Doing so allows us to measure the training time for each model and to compare the performance trade-off to the off-line training time. 
Since we are training relatively small networks, GPU training does not speed up the process in most cases, as the overhead of distributing across the cores is too large. 
All networks are trained on the Black--Scholes and Heston pricing problems as well as on the two implied volatility problems. 
The empirical results for each problem are discussed below.


\subsubsection{Black--Scholes model}

\begin{figure}[h!]
	\centering
	\includegraphics[width=\linewidth]{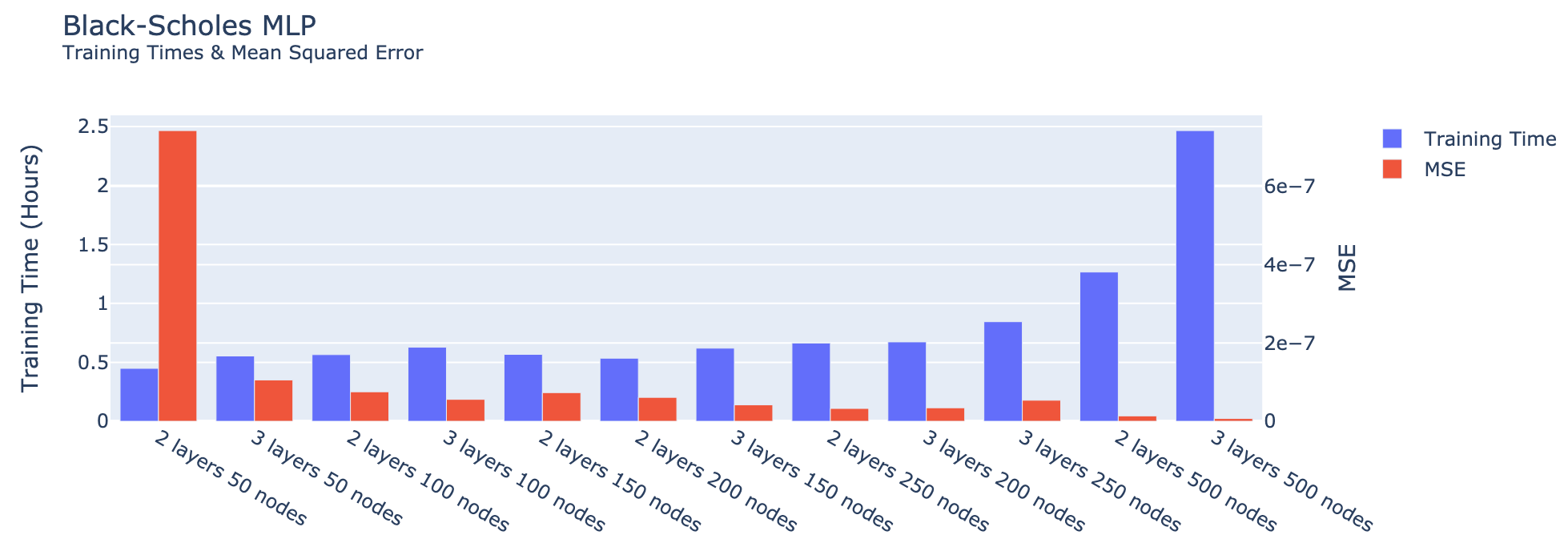}
	\caption{A comparison of the twelve networks from Table \ref{table:mlp_set} including MSE on the test set (red) and training time (blue) for the Black--Scholes problem.}
	\label{fig:bs_mlp_mse_training_set}
\end{figure}

Figure \ref{fig:bs_mlp_mse_training_set} visualizes the training time alongside the MSE for each network on the test set. 
We can see a clear relationship between the size of the network, which is increasing on the horizontal axis, the reduction in MSE and the increase in training time. 
Indeed, increasing the network size decreases the MSE (red) and hence improves the accuracy of the estimation, while the cost associated with this improvement is visible through the increase in training time (blue). 

\begin{figure}[h!]
	\centering
	\includegraphics[width=\linewidth]{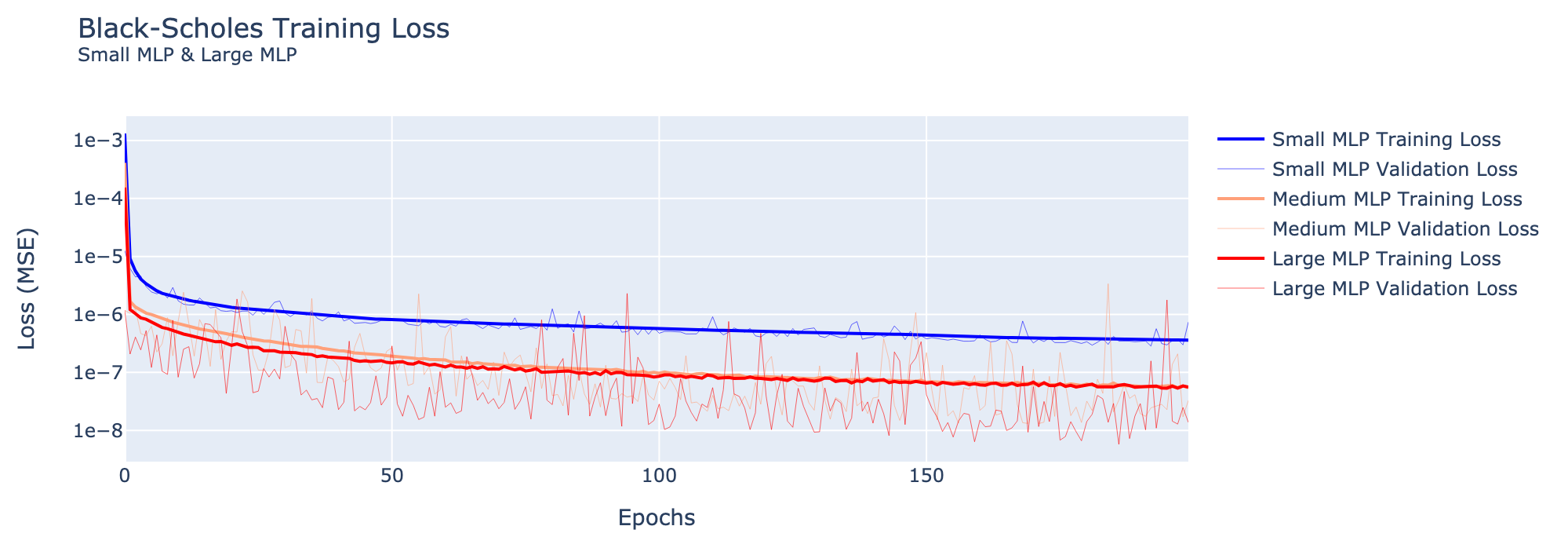}
	\caption{The training and validation losses for the three MLP networks. The small, medium and large networks contain $2,\!851$, $64,\!251$ and $504,\!001$ parameters respectively.}
	\label{fig:black_scholes_mlp_mse_training}
\end{figure}

Interestingly, the performance increase of a larger network is not immediately visible during training. 
Figure \ref{fig:black_scholes_mlp_mse_training} visualizes the training and validation losses over the $200$ training epochs for the $2$ layer networks with $50$ nodes (small), $250$ nodes (medium) and $500$ nodes (large). 
Notice that even though the largest network is almost $5$ times larger than the medium network and performs roughly $3.5$ times as well as the medium network, the training loss is almost identical for both networks. 
This is an indication that losses during training do not represent actual performance on unseen data. 
Additionally, the validation loss for the medium and large networks is lower than the training loss. 
This indicates a possible performance improvement for both networks if training time is extended for more epochs, which aligns with the results in \citet{pricing_options_implied_vol}. 
Since the Black--Scholes problem has an analytical solution, we expect to be able to approximate the solution using neural networks to any degree of precision, when choosing an appropriately sized network along with many training cycles. 
The results from this section support this hypothesis. 
We select the $3$ layer $50$ node network to compare against the highway and DGM networks discussed later, as a good trade-off between accuracy and training time.
Moreover, for the sake of completeness, we list the configuration of each network in Table \ref{table:mlp_statistics_twelve_models_bs}, including training time and MSE; here, training time is measured in fractions of an hour. 

\begin{table}[h!]
\centering
\begin{tabular}{ c|c|c|r|c|c } 
\hline
	Model & Layers & Nodes & Parameters & Training Time (H) & MSE \\ 
	\hline
	1 & $2$ & $50$ & $2,\!851$ & $0.52$ & $2.3 \cdot 10^{-7}$ \\
	2 & $2$ & $100$ & $10,\!701$ & $0.57$ & $7.5 \cdot 10^{-8}$ \\
	3 & $2$ & $150$ & $23,\!551$ & $0.57$ & $7.3 \cdot 10^{-8}$ \\
	4 & $2$ & $200$ & $41,\!401$ & $0.54$ & $6.1 \cdot 10^{-8}$ \\
	5 & $2$ & $250$ & $64,\!251$ & $0.66$ & $3.3 \cdot 10^{-8}$ \\
	6 & $2$ & $500$ & $253,\!501$ & $1.27$ & $1.4 \cdot 10^{-8}$ \\
	7 & $3$ & $50$ & $5,\!401$ & $0.55$ & $1.1 \cdot 10^{-7}$ \\
	8 & $3$ & $100$ & $20,\!801$ & $0.63$ & $5.6 \cdot 10^{-8}$ \\
	9 & $3$ & $150$ & $46,\!201$ & $0.62$ & $4.2 \cdot 10^{-8}$ \\
	10 & $3$ & $200$ & $81,\!601$ & $0.67$ & $3.4 \cdot 10^{-8}$ \\
	11 & $3$ & $250$ & $127,\!001$ & $0.84$ & $5.4 \cdot 10^{-8}$ \\
	\rowcolor{table_yellow}
	12 & $3$ & $500$ & $504,\!001$ & $2.47$ & $7.0 \cdot 10^{-9}$ \\
	MLP \cite{jph-report} & $6$ & $200$ & $202,\!201$ & - & $1.3 \cdot 10^{-7}$ \\
\hline
\end{tabular}
\caption{An overview of the MLP configurations alongside training time and MSE for the Black--Scholes problem. The best performing network (in MSE) is highlighted.}
\label{table:mlp_statistics_twelve_models_bs}
\end{table}


\subsubsection{Heston model}

\begin{figure}[h!]
	\centering
	\includegraphics[width=\linewidth]{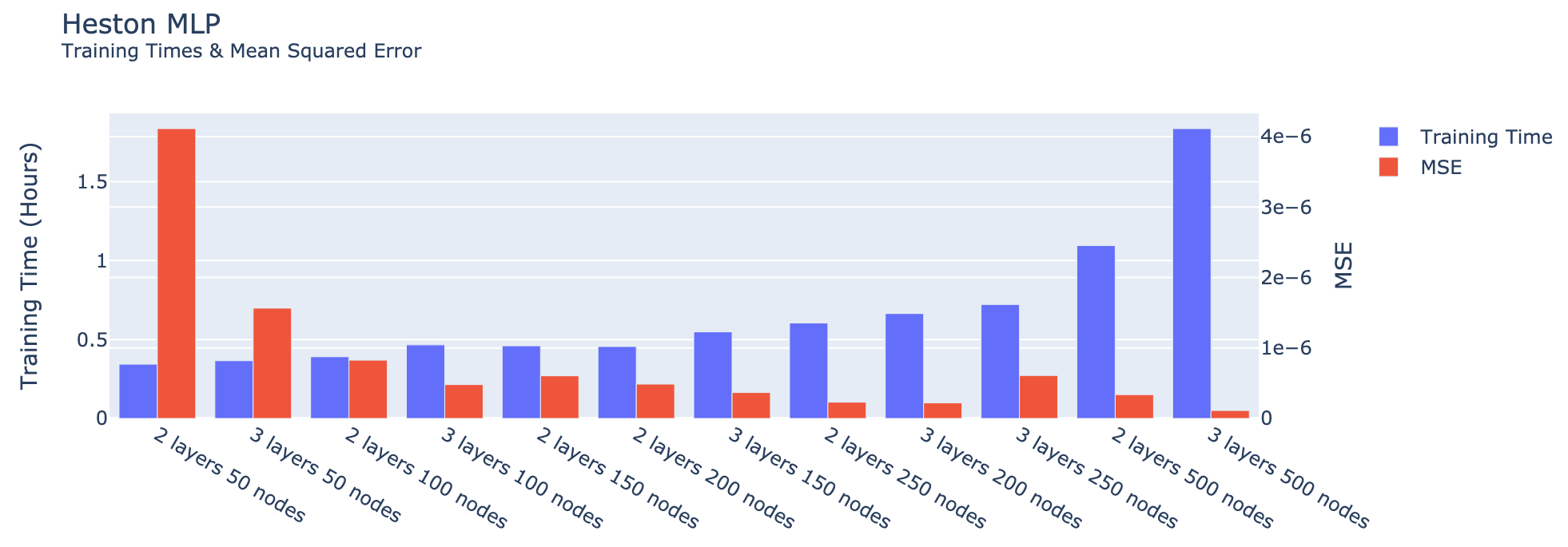}
	\caption{A comparison of the twelve networks from Table \ref{table:mlp_set} including MSE on the test set (red) and training time (blue) for the Heston problem.}
	\label{fig:heston_mlp_training_mse}
\end{figure}

The results of the various MLPs trained and evaluated on the Heston pricing problem are presented in Figure \ref{fig:heston_mlp_training_mse}. 
Therein we observe that the results for the Heston pricing problem using the MLP network architectures align qualitatively with those for the Black--Scholes problem. 
Since both problems are very similar in nature and attempt to approximate the same quantity, the similarity in the results is rather expected. 

Indeed, the results from both the Black--Scholes as well as the Heston pricing problem exhibit, aside from the magnitude of the MSEs, many similarities; compare Figures \ref{fig:bs_mlp_mse_training_set} and \ref{fig:heston_mlp_training_mse}. 
The MSEs and training times for the Heston model can be found in Table \ref{table:mlp_statistics_twelve_models_heston}.
The difference in MSEs is not surprising, as the Heston model has 8 parameters compared to the 4 parameters in the Black--Scholes model.

\begin{table}[h!]
\centering
\begin{tabular}{ c|c|c|r|c|c } 
\hline
	Model & Layers & Nodes & Parameters & Training Time (H) & MSE \\ 
	\hline
	1 & $2$ & $50$ & $3,\!051$ & $0.34$ & $4.1 \cdot 10^{-6}$ \\
	2 & $2$ & $100$ & $11,\!101$ & $0.39$ & $8.3 \cdot 10^{-7}$ \\
	3 & $2$ & $150$ & $24,\!151$ & $0.46$ & $6.0 \cdot 10^{-7}$ \\
	4 & $2$ & $200$ & $42,\!201$ & $0.46$ & $4.9 \cdot 10^{-7}$ \\
	5 & $2$ & $250$ & $65,\!251$ & $0.60$ & $2.3 \cdot 10^{-7}$ \\
	6 & $2$ & $500$ & $255,\!501$ & $1.09$ & $3.4 \cdot 10^{-7}$ \\
	7 & $3$ & $50$ & $5,\!601$ & $0.37$ & $1.6 \cdot 10^{-6}$ \\
	8 & $3$ & $100$ & $21,\!201$ & $0.47$ & $4.8 \cdot 10^{-7}$ \\
	9 & $3$ & $150$ & $46,\!801$ & $0.55$ & $3.7 \cdot 10^{-7}$ \\
	10 & $3$ & $200$ & $82,\!401$ & $0.66$ & $2.2 \cdot 10^{-7}$ \\
	11 & $3$ & $250$ & $128,\!001$ & $0.72$ & $6.1 \cdot 10^{-7}$ \\
	12 & $3$ & $500$ & $506,\!001$ & $1.84$ & $1.1 \cdot 10^{-7}$ \\
	\rowcolor{table_yellow}
	MLP \cite{pricing_options_implied_vol} & $4$ & $400$ & $646,\!006$ & - & $1.7 \cdot 10^{-8}$ \\
\hline
\end{tabular}
\caption{An overview of the MLP configurations alongside training time and MSE for the Heston pricing problem. The best performing network (in MSE) is highlighted.}
\label{table:mlp_statistics_twelve_models_heston}
\end{table}

In Table \ref{table:mlp_statistics_twelve_models_heston}, we observe that the MLP from \cite{pricing_options_implied_vol} performs $10$ times better than the best performing MLP that we trained. We trained each network for $200$ epochs, whereas the authors of \cite{pricing_options_implied_vol} train for $3000$ epochs. 
Since the architecture of the networks is identical, this indicates once again that longer training on this problem improves performance, similar to what we observed for the Black--Scholes pricing problem.


\subsubsection{Implied volatility}\label{subsec:mlp_iv}

\begin{figure}[h!]
	\centering
	\includegraphics[width=\linewidth]{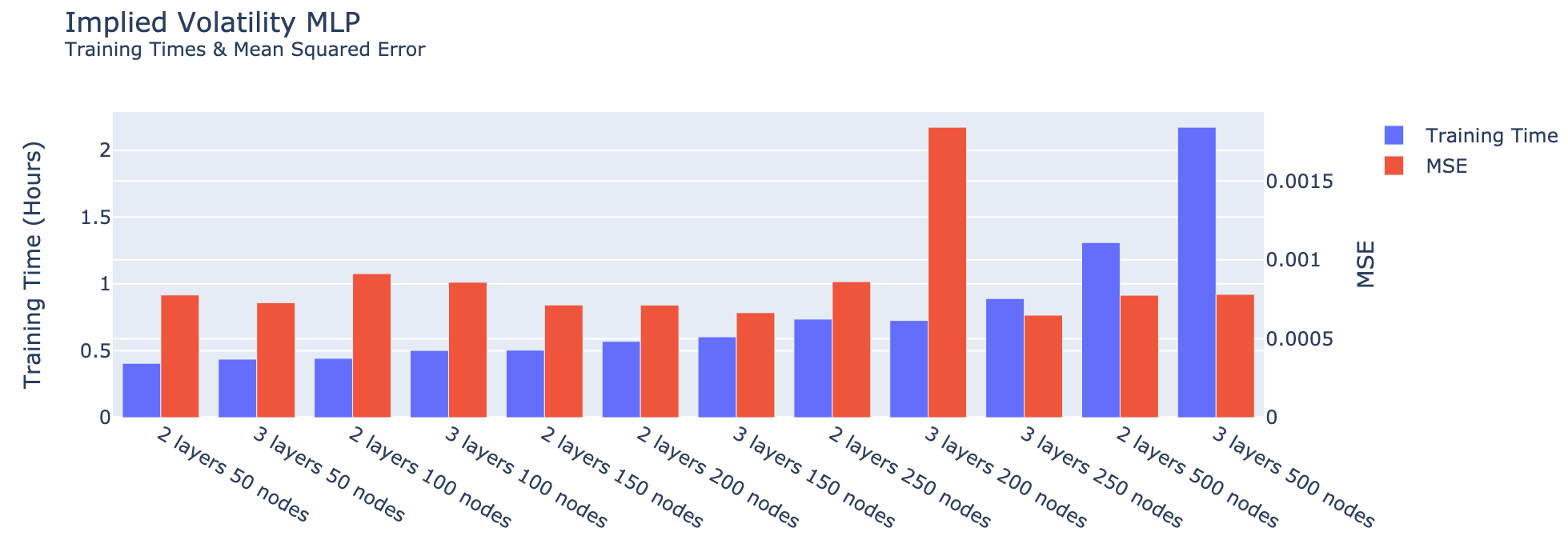}
	\caption{A comparison of the twelve networks from Table \ref{table:mlp_set} including MSE on the test set (red) and training time (blue) for the implied volatility problem.}
	\label{fig:iv_mlp_mse_training_set}
\end{figure}

The main observation for the implied volatility problem is that, contrary to the Black--Scholes and Heston problems, we cannot find evidence of larger networks performing better for this (inverse) problem. 
Indeed, the smallest network we train consists of $2$ layers and $50$ nodes per layer, containing a total of $2,\!851$ parameters, and performs just as well after $200$ epochs with identical configuration as the largest network with $3$ layers and $500$ nodes, containing $504,\!001$ parameters. 
Both networks have an MSE on the test set of $7.7 \cdot 10^{-4}$, while the small network took only $0.4$ hours to train compared to the $2.17$ hours for the largest network. 
All networks show comparable performance, as can be seen in Figure \ref{fig:iv_mlp_mse_training_set}.

\begin{figure}[h!]
	\centering
	\includegraphics[width=\linewidth]{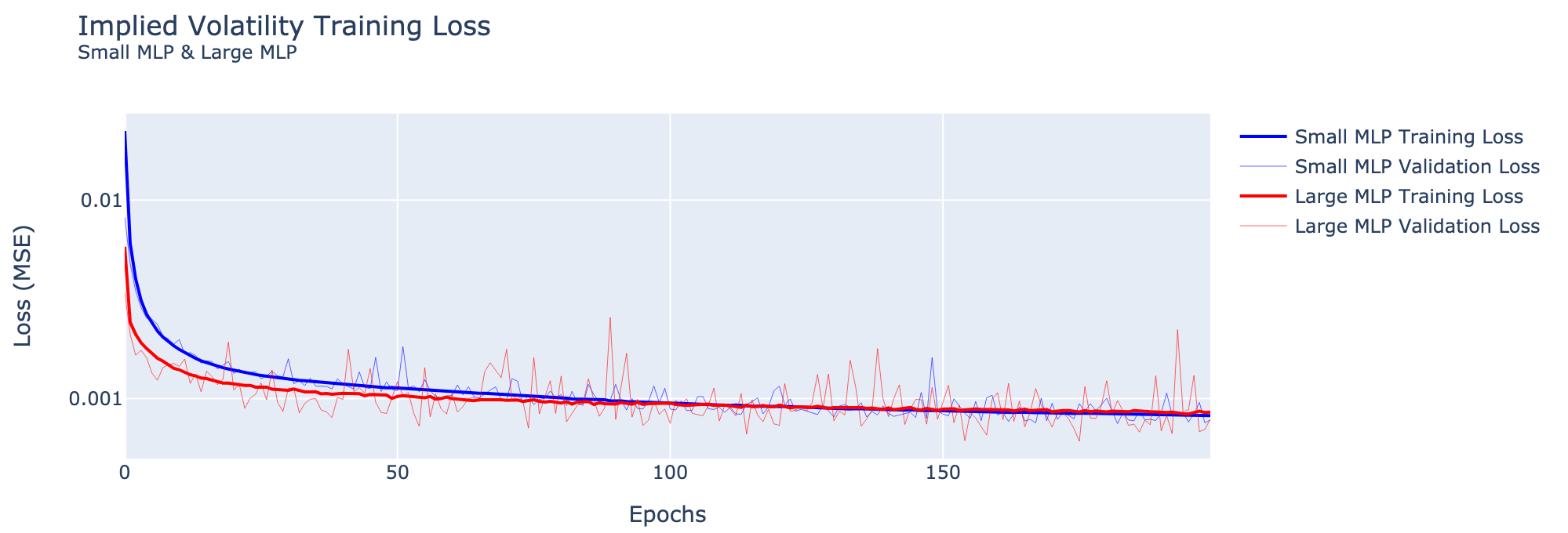}
	\caption{The training and validation loss of the smallest network compared to the largest network on the implied volatility data set.}
	\label{fig:iv_mlp_mse_small_large_model}
\end{figure}

The identical performance of the networks during training is visualized in Figure \ref{fig:iv_mlp_mse_small_large_model}. 
The results we find align with the errors found in \cite{jph-report}, in which the best performing MLP network containing $202,\!201$ parameters performed worse on the test data than the considerably smaller networks from Figure \ref{fig:iv_mlp_mse_training_set}.
Table \ref{table:mlp_statistics_twelve_models_iv} lists all the networks along with the training times, number of parameters and MSE for the implied volatility problem.	
The results indicate that the networks are dealing with convergence issues that prevent them from further optimization, regardless of their size and architecture. 

Next, we consider the transformed implied volatility dataset, where we aim to solve the steep gradient problems possibly causing the convergence issues by introducing a transformation from the scaled call price to the scaled time value of the option, explained in Subsection \ref{subsec:implied_volatility_problem_description}.

\begin{table}[h!]
\centering
\begin{tabular}{ c|c|c|r|c|c } 
\hline
	Model & Layers & Nodes & Parameters & Training Time (H) & MSE \\ 
	\hline
	1 & $2$ & $50$ & $2,\!851$ & $0.40$ & $7.7 \cdot 10^{-4}$ \\
	2 & $2$ & $100$ & $10,\!701$ & $0.44$ & $9.1 \cdot 10^{-4}$ \\
	3 & $2$ & $150$ & $23,\!551$ & $0.50$ & $7.1 \cdot 10^{-4}$ \\
	4 & $2$ & $200$ & $41,\!401$ & $0.57$ & $7.1 \cdot 10^{-4}$ \\
	5 & $2$ & $250$ & $64,\!251$ & $0.74$ & $8.6 \cdot 10^{-4}$ \\
	6 & $2$ & $500$ & $253,\!501$ & $1.31$ & $7.8 \cdot 10^{-4}$ \\
	7 & $3$ & $50$ & $5,\!401$ & $0.44$ & $7.3 \cdot 10^{-4}$ \\
	8 & $3$ & $100$ & $20,\!801$ & $0.50$ & $8.5 \cdot 10^{-4}$ \\
	\rowcolor{table_yellow}
	9 & $3$ & $150$ & $46,\!201$ & $0.60$ & $6.6 \cdot 10^{-4}$ \\
	10 & $3$ & $200$ & $81,\!601$ & $0.73$ & $1.8 \cdot 10^{-3}$ \\
	11 & $3$ & $250$ & $127,\!001$ & $0.94$ & $7.6 \cdot 10^{-4}$ \\
	12 & $3$ & $500$ & $504,\!001$ & $2.17$ & $7.7 \cdot 10^{-4}$ \\
\hline
\end{tabular}
\caption{An overview of the MLP configurations alongside training time and MSE for the implied volatility problem. The best performing network (in MSE) is highlighted.}
\label{table:mlp_statistics_twelve_models_iv}
\end{table}

\begin{figure}[h!]
	\centering
	\includegraphics[width=\linewidth]{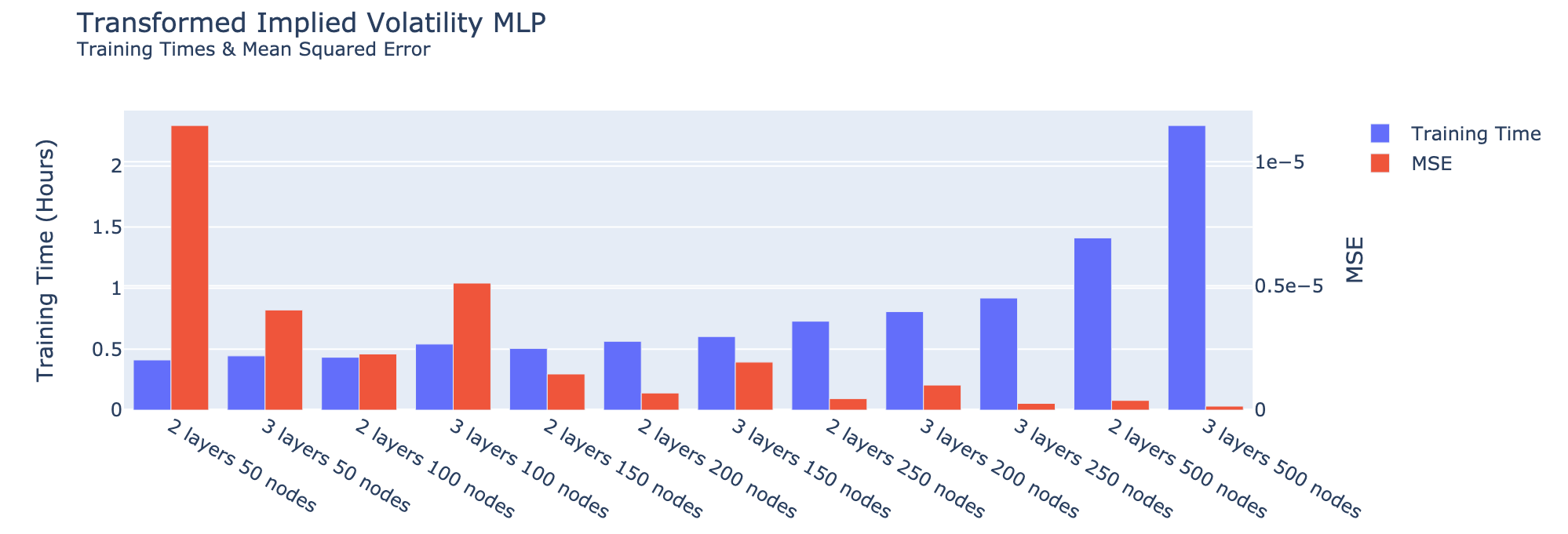}
	\caption{A comparison of the twelve networks from Table \ref{table:mlp_set} including MSE on the test set (red) and training time (blue) for the transformed implied volatility problem.}
	\label{fig:transformed_iv_mlp}
\end{figure}

Figure \ref{fig:transformed_iv_mlp} visualizes the performance of the MLPs on the transformed implied volatility problem. 
We notice that on the transformed problem, the MSE is decreasing as a function of the network size, similarly to what we found for the option pricing problems before. 
The reduction in MSE is not as consistent as in the previous problems, indicating difficult convergence during training, possibly still due to a steep gradient not being entirely removed by the transformation. 
The results are considerably better than for the `default' implied volatility problem, using only a simple transformation. 
Table \ref{table:mlp_statistics_twelve_models_transformed_iv} shows the results of all MLPs alongside their parameters and training times.

\begin{table}[h!]
\centering
\begin{tabular}{ c|c|c|r|c|c } 
\hline
	Model & Layers & Nodes & Parameters & Training Time (H) & MSE \\ 
	\hline
	1 & $2$ & $50$ & $2,\!851$ & $0.41$ & $1.1 \cdot 10^{-5}$ \\
	2 & $2$ & $100$ & $10,\!701$ & $0.43$ & $2.3 \cdot 10^{-6}$ \\
	3 & $2$ & $150$ & $23,\!551$ & $0.51$ & $1.5 \cdot 10^{-6}$ \\
	4 & $2$ & $200$ & $41,\!401$ & $0.56$ & $1.9 \cdot 10^{-6}$ \\
	5 & $2$ & $250$ & $64,\!251$ & $0.73$ & $6.9 \cdot 10^{-6}$ \\
	6 & $2$ & $500$ & $253,\!501$ & $1.41$ & $3.9 \cdot 10^{-7}$ \\
	7 & $3$ & $50$ & $5,\!401$ & $0.44$ & $4.0 \cdot 10^{-6}$ \\
	8 & $3$ & $100$ & $20,\!801$ & $0.54$ & $5.1 \cdot 10^{-6}$ \\
	9 & $3$ & $150$ & $46,\!201$ & $0.60$ & $1.9 \cdot 10^{-6}$ \\
	10 & $3$ & $200$ & $81,\!601$ & $0.81$ & $1.0 \cdot 10^{-6}$ \\
	11 & $3$ & $250$ & $127,\!001$ & $0.92$ & $2.7 \cdot 10^{-7}$ \\
	\rowcolor{table_yellow}
	12 & $3$ & $500$ & $504,001$ & $2.33$ & $1.6 \cdot 10^{-7}$ \\
\hline
\end{tabular}
\caption{An overview of the MLP configurations alongside training time and MSE for the transformed implied volatility problem. The best performing network (in MSE) is highlighted.}
\label{table:mlp_statistics_twelve_models_transformed_iv}
\end{table}

In order to quantify the practical impact of the log-transformation we compare the performance of identical architectures on the default and transformed implied volatility datasets.
Using the 2-layer MLP with 150 nodes per layer, the default formulation (Table \ref{table:mlp_statistics_twelve_models_iv}) yields a test MSE of $7.1 \times 10^{-4}$, whereas the transformed formulation (Table \ref{table:mlp_statistics_twelve_models_transformed_iv}) reduces the error to $1.5 \times 10^{-6}$. This improvement by almost three orders of magnitude is achieved without changing the architecture, hyperparameters, or training regime. 
The substantial gain in accuracy confirms empirically that the time-value and log-transform significantly improve numerical conditioning and optimization stability, consistent with the gradient analysis in Section \ref{subsec:implied_volatility_problem_description}.


\subsection{Highway networks}
\label{subsec:highway_network_analysis}

In this section, we train variants of the highway networks discussed in Section \ref{subsec:highway_networks} with approximately equal configurations. 
More specifically, each network contains $3$ layers each with $50$ nodes. 
We train relatively small networks to compare across various architectures, in order to keep computational times feasible. 
Table \ref{table:highway_network_types} lists each highway network configuration.

\begin{table}[h!]
\centering
\begin{tabular}{ c|c|c|c } 
\hline
 & Generalized Highway & Highway & Residual \\ 
\hline
	Layers & $3$ & $3$ & $3$ \\
	Nodes per layer & $50$ & $50$ & $50$ \\
	Total parameters & $23,\!251$ & $15,\!601$ & $7,\!951$ \\
	Initializer & Glorot Normal & Glorot Normal & Glorot Normal \\
	Carry/Transform activation function & $\tanh$ & $\tanh$ & - \\ 
	Activation function & $\tanh$ & $\tanh$ & $\tanh$ \\
	Loss function & MSE & MSE & MSE \\
	Learning rate & $10^{-5}$ & $10^{-5}$ & $10^{-5}$ \\
	Batch size & $64$ & $64$ & $64$ \\
\hline
\end{tabular}
\caption{Overview of highway network configurations.}
\label{table:highway_network_types}
\end{table}


\subsubsection{Black--Scholes model}

\begin{figure}[h!]
	\centering
	\includegraphics[width=0.8\linewidth]{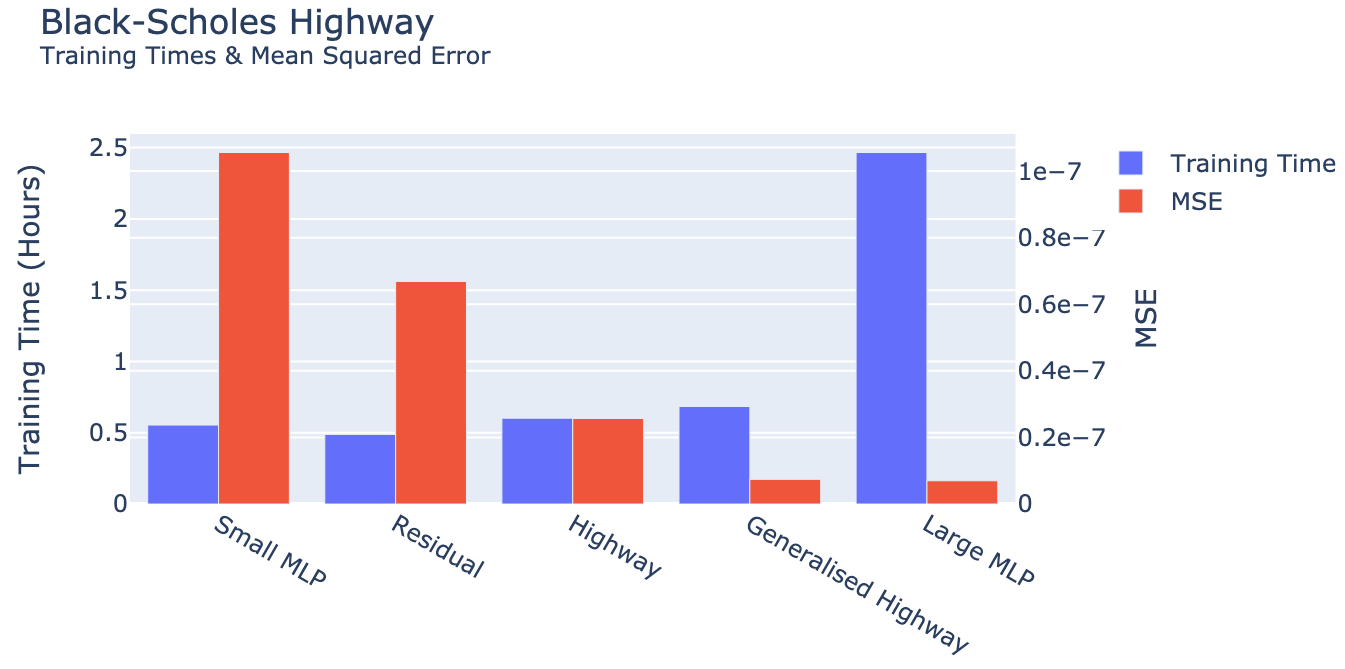}
	\caption{A comparison of highway networks from Table \ref{table:highway_network_types} including the smallest and largest MLP with the MSE on the test set (red) and training time (blue) for the Black--Scholes problem. The small MLP contains $3$ layers with $50$ nodes, and the large MLP $3$ layers with $500$ nodes. The results are ordered by the total number of parameters.}
	\label{fig:bs_highway_training_set}
\end{figure}

\begin{figure}[h!]
	\centering
	\includegraphics[width=0.8\linewidth]{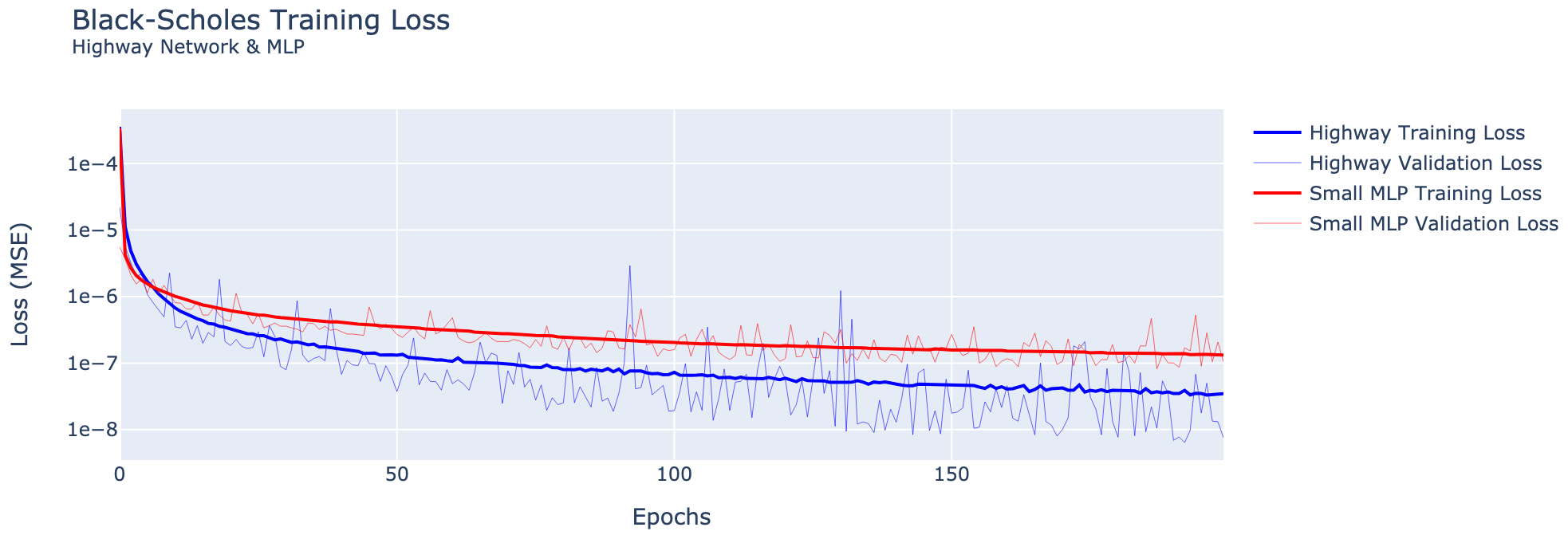}
	\caption{The training and validation loss over $200$ epochs. The small MLP consisting of $3$ layers and $50$ nodes ($2,851$ parameters) in red, compared to the $3$ layer $50$ node generalized highway network ($23,251$ parameters) in blue.}
	\label{fig:bs_highway_training_validation_loss}
\end{figure}

Figure \ref{fig:bs_highway_training_set} visualizes the training times (blue) and MSE on the test set (red) for each highway network. 
Both the highway and generalized highway networks perform better than their MLP counterpart with an equal number of nodes. 
More impressive, however, is the fact that both networks required barely more training time than the small MLP, but scored comparably to the large MLP on the test set. 
This means that the highway architecture improves convergence on the Black--Scholes problem significantly. 
The difference in convergence speed can also be seen during training, as shown in Figure \ref{fig:bs_highway_training_validation_loss}. 
In this case, training losses are a good indicator for performance on the test set, likely because the performance difference for both networks is so large. 
The loss during training is roughly one order of magnitude smaller for the generalized highway model compared to the MLP, whereas the training time is only $23\%$ longer; \textit{cf.} Table \ref{table:highway_performance_overview}.  

The residual layer is closest to the classical dense layer. 
The only difference is the carry of the input $\mathbf{x}$ to the next layer, \textit{cf.} \eqref{eq:residual_layer_operation}. 
Surprisingly, only carrying the input vector to the next layer decreases the MSE on the test set by $33\%$ from $1.1 \cdot 10^{-7}$ to $6.7 \cdot 10^{-8}$. 
Since only one additional operation is added in each layer, the training time is almost identical. 
\citet{MR3874585}, in the context of solutions for PDEs, argue that including the input vector in many operations allows the network to make `sharper turns', resulting in a more flexible network. 
We see that this feature becomes particularly useful also when approximating the Black--Scholes price.

The difference in performance between the highway network and the residual network is also significant. 
The incorporation of one extra weight matrix and one additional non-linear function multiplied with the input vector $\mathbf{x}$ in each layer, \textit{cf.} \eqref{eq:simplified_highway_layer}, results in an MSE reduction of $61\%$ from the residual network's $6.7 \cdot 10^{-8}$ to the highway network's $2.6 \cdot 10^{-8}$. 
Adding one additional weight matrix $W_C$ and non-linearity $C$, yielding the generalized highway layer, improves MSE performance by another $61\%$ compared to the highway layer. 
An overview of the error reductions, training times and operations in each layer are given in Table \ref{table:highway_performance_overview}. 

\begin{table}[h!]
\centering
\begin{tabular}{ c|c|c|c|c } 
	\hline
	 & Time (H) & MSE & Reduction ($\%$) & Layer operation \\ 
	 \hline
	 MLP & $0.55$ & $1.1 \cdot 10^{-7}$ & - & $H$ \\
	 Residual & $0.55$ & $6.7 \cdot 10^{-8}$ & $33\%$ & $H + \mathbf{x}$ \\
	 Highway & $0.60$ & $2.8 \cdot 10^{-8}$ & $61\%$ & $H \cdot T + \mathbf{x} \cdot (1 - T)$ \\
	 Generalized Highway & $0.68$ & $7.4 \cdot 10^{-9}$ & $73\%$ & $H \cdot T + \mathbf{x} \cdot C$ \\ 
	\hline
\end{tabular}
\caption{An overview of the highway network performance and error reductions for the MLP and highway network architectures on the Black--Scholes pricing problem. The reductions in the fourth column are relative to the previous row.}
\label{table:highway_performance_overview}
\end{table}

Finally, we notice that a $23\%$ increase in training time from the small MLP to the generalized highway network leads to a reduction of the MSE by $99\%$.  
When comparing the results obtained for the highway networks to the results in \citet[Table~6]{pricing_options_implied_vol}, in which a large MLP is trained and evaluated on identical data, we see that our small generalized highway network already outperforms the MLP. 
In fact, as shown in Table \ref{table:statistics_bs_dgm}, the generalized highway network improves the MSE by $9.8\%$ while reducing the amount of parameters by more than $96\%$ compared to the MLP from \cite{pricing_options_implied_vol}.


\subsubsection{Heston model}

\begin{figure}[h!]
	\centering
	\includegraphics[width=0.8\linewidth]{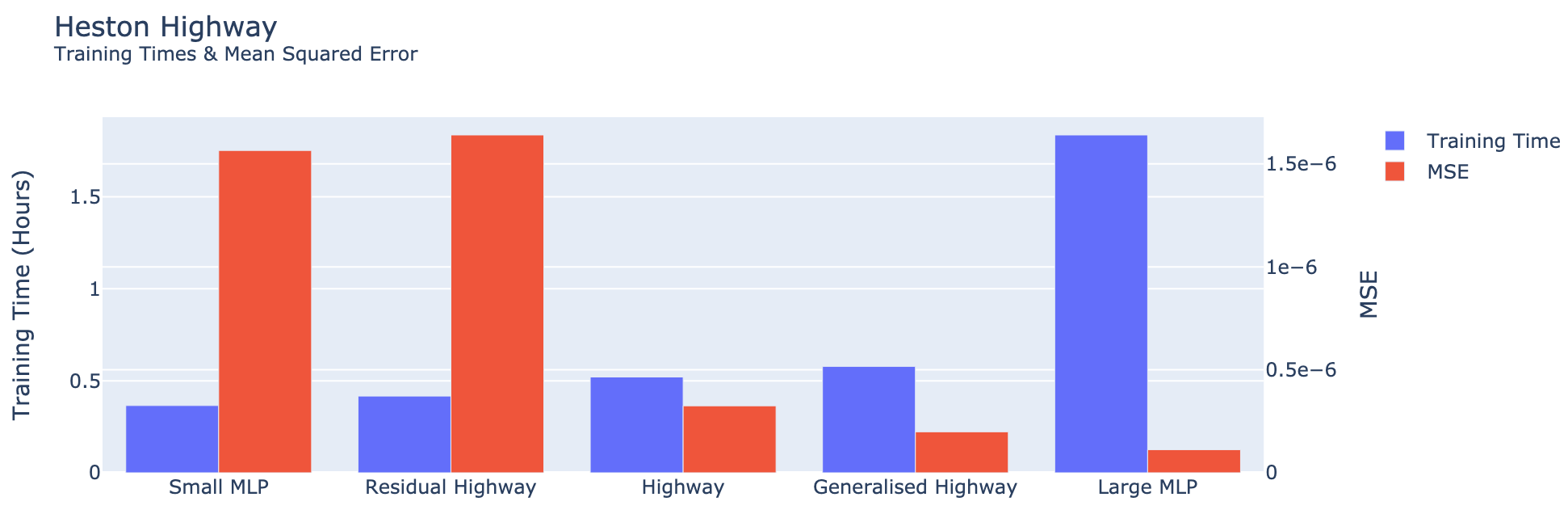}
	\caption{A comparison of highway networks from Table \ref{table:highway_network_types} including the smallest and largest MLP with the MSE on the test set (red) and training time (blue) on the Heston pricing problem. The small MLP contains $3$ layers with $50$ nodes, and the large MLP contains $3$ layers with $500$ nodes.}
	\label{fig:heston_highway_training_set}
\end{figure}

Figure \ref{fig:heston_highway_training_set} visualizes the performance of the highway networks for the Heston pricing problem. 
We observe again a similar behaviour as in the Black--Scholes pricing problem; compare with Figure \ref{fig:bs_highway_training_set}. 
Indeed, the highway and generalized highway networks perform comparably to the large MLP in terms of MSE, while requiring the training time of the small MLP.
A major difference is the performance of the residual network, which reduced the MSE by $33\%$ compared to the MLP on the Black--Scholes problem; \textit{cf.}  Table \ref{table:highway_performance_overview}. 
On the Heston pricing problem, we do not observe an increase in performance. 
A possible explanation for this could be that the Heston pricing problem might be too involved for the residual architecture to make an improvement.

\begin{table}[h!]
\centering
\begin{tabular}{ c|c|c|c|c } 
\hline
	 & Time (H) & MSE & Reduction ($\%$) & Layer Operation \\ 
	 \hline
	 MLP & $0.37$ & $1.6 \cdot 10^{-6}$ & - & $H$ \\
	 Residual & $0.42$ & $1.7 \cdot 10^{-6}$ & $-5\%$ & $H + \mathbf{x}$ \\
	 Highway & $0.58$ & $3.3 \cdot 10^{-7}$ & $80\%$ & $H \cdot T + \mathbf{x} \cdot (1 - T)$ \\
	 Generalized Highway & $0.68$ & $2.0 \cdot 10^{-7}$ & $39\%$ & $H \cdot T + \mathbf{x} \cdot C$ \\ 
\hline
\end{tabular}
\caption{An overview of the highway network performance and error reductions for the MLP and highway network architectures on the Heston pricing problem. The
reductions in the fourth column are relative to the previous row.}
\label{table:highway_performance_overview_heston}
\end{table}

The highway and generalized highway networks do, in fact, significantly improve performance compared to the MLP networks, with the generalized highway network performing almost as well as the largest MLP, while requiring only $31\%$ of the computational time. 
Table \ref{table:highway_performance_overview_heston} lists the reductions that each highway variation contributes to the MSE. 
We notice that the improvement of the generalized highway network on the Heston problem is less significant than on the Black--Scholes problem, where the generalized highway network was the best performing network. 
In Section \ref{subsec:dgm_network_analysis} we compare the highway networks to their DGM counterpart on the Heston pricing problem, and consider equally sized networks in terms of parameters.


\subsubsection{Implied volatility}

\begin{figure}[h!]
	\centering
	\includegraphics[width=0.8\linewidth]{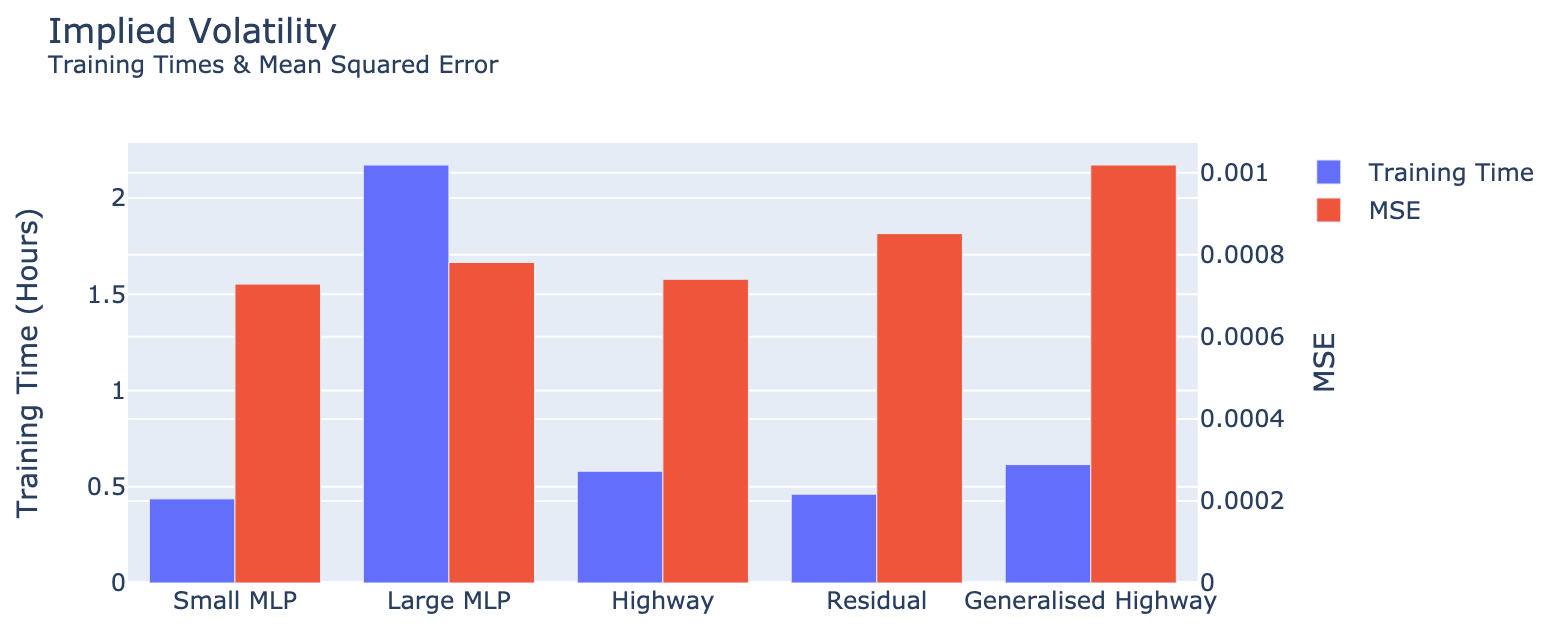}
	\caption{A comparison of the highway networks from Table \ref{table:highway_network_types} including the smallest and largest MLP with the MSE on the test set (red) and training time (blue) on the implied volatility problem. The small MLP contains $3$ layers with $50$ nodes and the large MLP contains $3$ layers with $500$ nodes.}
	\label{fig:iv_highway_training_set}
\end{figure}

In the previous subsections, we found significant reductions in MSE on the test set when comparing the highway networks to the MLPs for the Black--Scholes and the Heston pricing problems. 
In this subsection, we perform the same comparison for the implied volatility problems. 
The results are visualized in Figure \ref{fig:iv_highway_training_set}. 
In contrast to the Black--Scholes problem, in which the highway networks outperformed the MLPs, this is not the case for the implied volatility problem. 
None of the highway networks outperforms the MLPs on the test set. 
As in the MLP scenario, this indicates again that the steep gradient also poses a problem when optimizing the highway networks. 
In order to mitigate this issue, we consider the performance of the networks on the transformed implied volatility problem.

\begin{figure}[h!]
	\centering
	\includegraphics[width=0.8\linewidth]{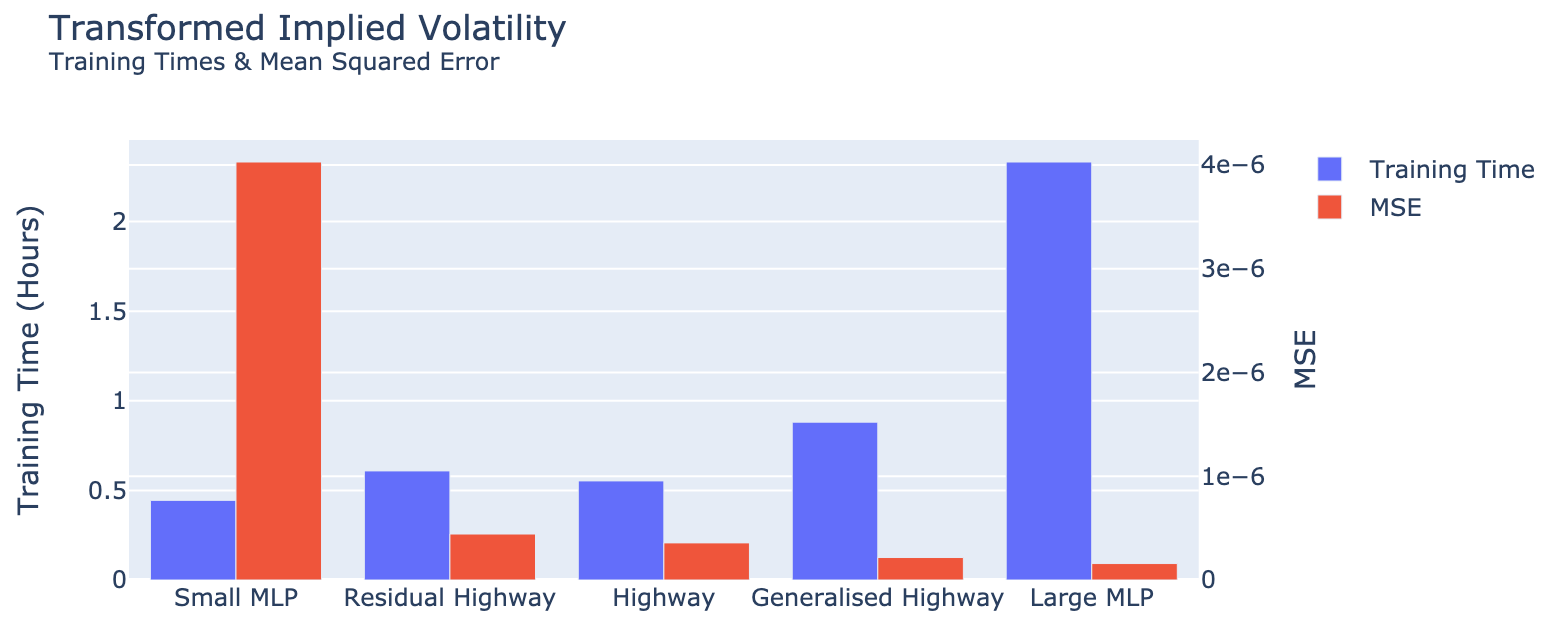}
	\caption{A comparison of the highway networks from Table \ref{table:highway_network_types} including the smallest and largest MLP with the MSE on the test set (red) and training time (blue) on the transformed implied volatility dataset. The small MLP contains $3$ layers with $50$ nodes and the large MLP contains $3$ layers with $500$ nodes.}
	\label{fig:transformed_iv_highway_training_set}
\end{figure}

In Figure \ref{fig:transformed_iv_highway_training_set} the performance of the highway networks on the transformed implied volatility dataset is shown. 
On this dataset, we find similar relative MSEs and computational times as for the Black--Scholes and Heston pricing problems. 
In fact, the generalized highway network is again the superior network relative to the training time. 
However, just like in the Heston pricing problem scenario, the large MLP scores lower in absolute MSE. 
This indicates that the network performance on the implied volatility problem is sensitive to larger network sizes. 
We expect to see performance improvements when training highway networks with more nodes and layers. 

These results clarify how much influence the network architecture has on the performance; consider, for example, the difference between the small MLP and the residual highway network. 
Recall that the operation of a residual network is simply the addition of the input vector to the layer; \textit{cf.} \eqref{eq:residual_layer_operation}. 
This operation reduces the MSE by roughly $89\%$, while adding only $38\%$ to the training time. 
We also notice that the training times for an identical network architecture can differ per dataset. 
In order to see this, consider the $14\%$ increase in training time from the small MLP to the residual network on the Heston problem (Table \ref{table:highway_performance_overview_heston}) and the $0\%$ increase on the Black--Scholes problem (Table \ref{table:highway_performance_overview}).


\subsection{DGM networks}
\label{subsec:dgm_network_analysis}

In this section, we train the DGM network and its variations presented in subsections \ref{subsec:DGM_network}--\ref{subsec:no_recurrence_dgm_network} and compare their performance to the highway and MLP networks. 
We train the DGM network with the configuration listed in Table \ref{table:dgm_configuration}. 

\begin{table}[h!]
\centering
\begin{tabular}{ c|c } 
\hline
 & DGM Network \\ 
\hline
Layers & $3$ \\
Nodes per layer & $50$ \\
Total parameters & $33,\!459$ \\
Activation function & $\tanh$ \\
Initialiser & Glorot Normal \\
Loss function & MSE \\
Learning rate & $10^{-5}$ \\
Batch size & 64 \\
\hline
\end{tabular}
\caption{Overview of DGM network configurations.}
\label{table:dgm_configuration}
\end{table} 


\subsubsection{Black--Scholes model}

\begin{figure}[h!]
	\centering
	\includegraphics[width=0.8\linewidth]{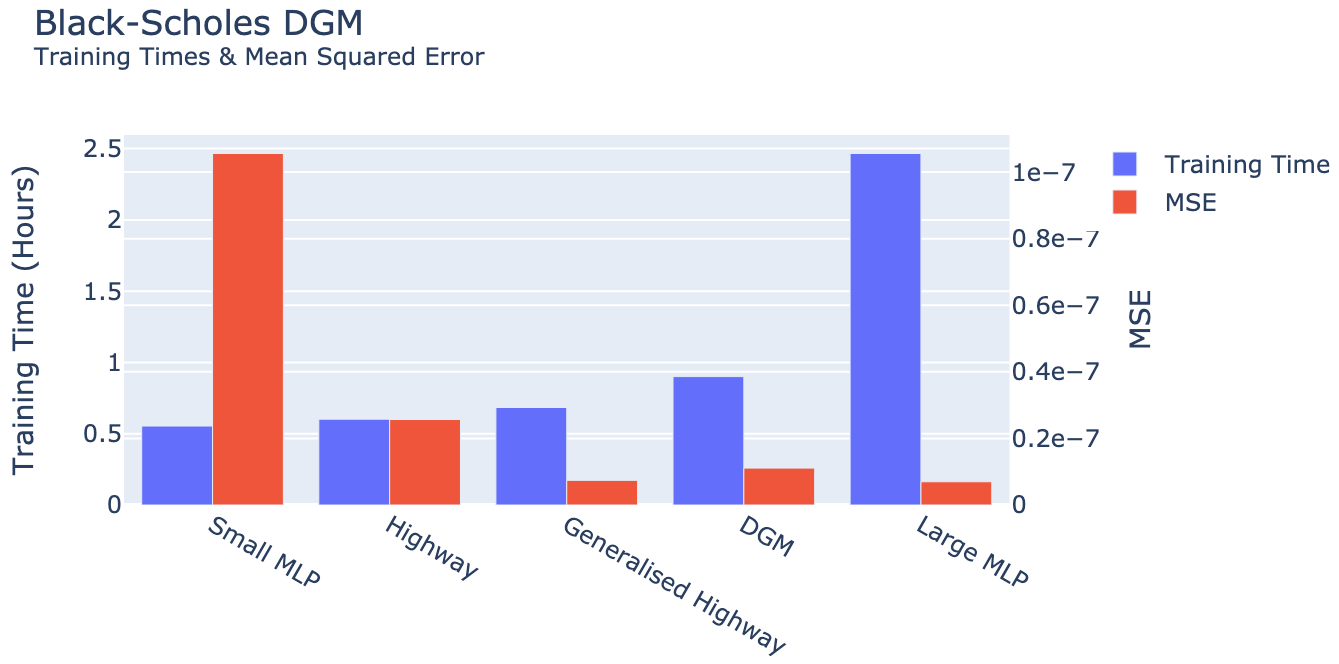}
	\caption{The MLP, highway and DGM networks compared in terms of MSE on the test set (red) and training time (blue) for the Black--Scholes model. The small MLP contains $3$ layers with $50$ nodes and the large MLP contains $3$ layers with $500$ nodes.}
	\label{fig:bs_dgm_mse_training}
\end{figure}

The results of the DGM network performance in the Black--Scholes problem are visualized in Figure \ref{fig:bs_dgm_mse_training}. 
We note that more operations must be computed inside a DGM network than in a highway network, resulting in slightly more parameters and hence in longer training time. 
This aligns with the numerical results. 

The main difference between the highway layer and the DGM layer is that the original input $\mathbf{x}$ is used in each DGM layer, whereas the highway layer only depends on the output of the previous layer.
The dependence on $\mathbf{x}$ introduces $4$ additional weight matrices per layer; see Figure \ref{fig:dgm_layer_schematic}. 
In Figure \ref{fig:bs_dgm_mse_training} no significant improvements due to this dependence compared to the generalized highway network can be observed.
Indeed, the generalized highway network has a smaller MSE and requires less computational time for training. 
Possibly, the Black--Scholes problem is not complex enough for the additional non-linearity and recurrence in the DGM layers to cause improvements. 

\begin{figure}[h!]
	\centering
	\includegraphics[width=0.85\linewidth]{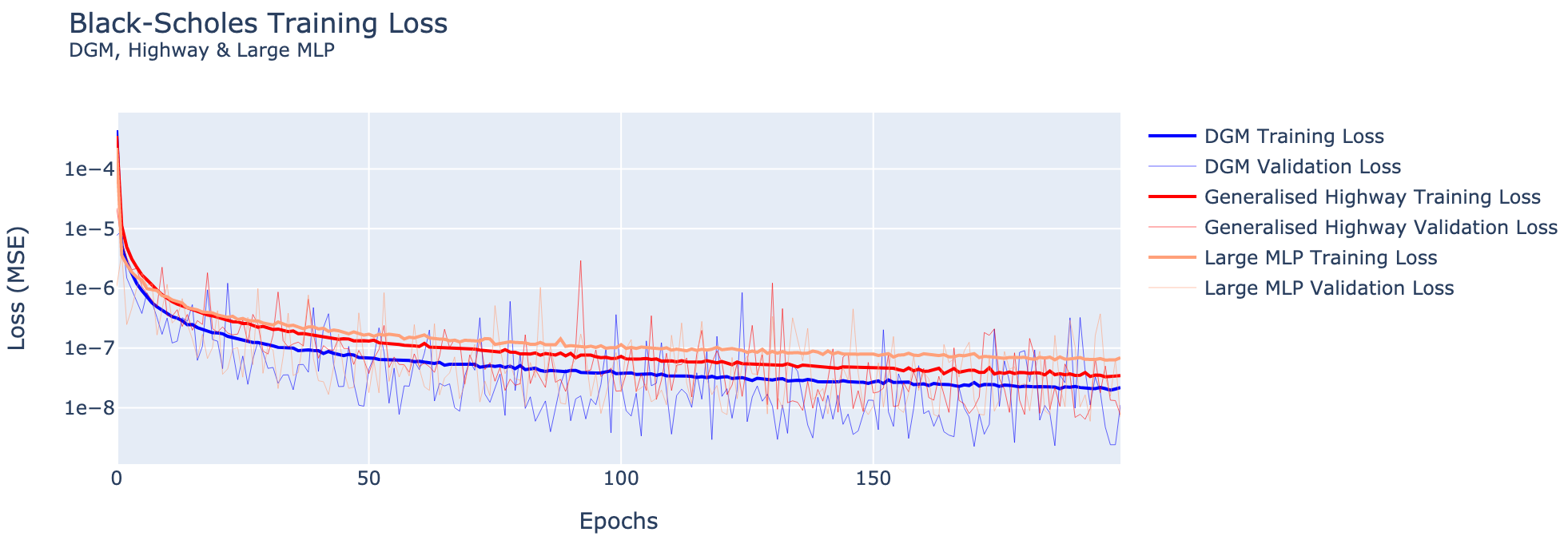}
	\caption{The MLP, highway and DGM networks compared during the training cycles.}
	\label{fig:bs_dgm_training_validation_loss}
\end{figure}

Figure \ref{fig:bs_dgm_training_validation_loss} reports the training and validation MSE for the DGM, generalised highway, and large MLP architectures applied to the Black–Scholes problem. 
The training and validation losses decrease concurrently for all models throughout the optimization and remain closely aligned, with no evidence of a systematic divergence between the two curves. 
This indicates that none of the architectures exhibit overfitting over the training horizon considered. 
Although the validation losses display higher variance, particularly at later epochs, their overall downward trend mirrors that of the training losses, which is consistent with stochastic simulation and Monte Carlo sampling effects rather than deteriorating generalization. 
Among the three architectures, DGM attains the lowest error levels, followed by the generalized highway network and the large MLP, reflecting differences in architectural inductive bias rather than generalization failure. Overall, the results demonstrate stable training dynamics and good generalization performance across all models.


\begin{figure}[h!]
	\centering
	\includegraphics[width=0.8\linewidth]{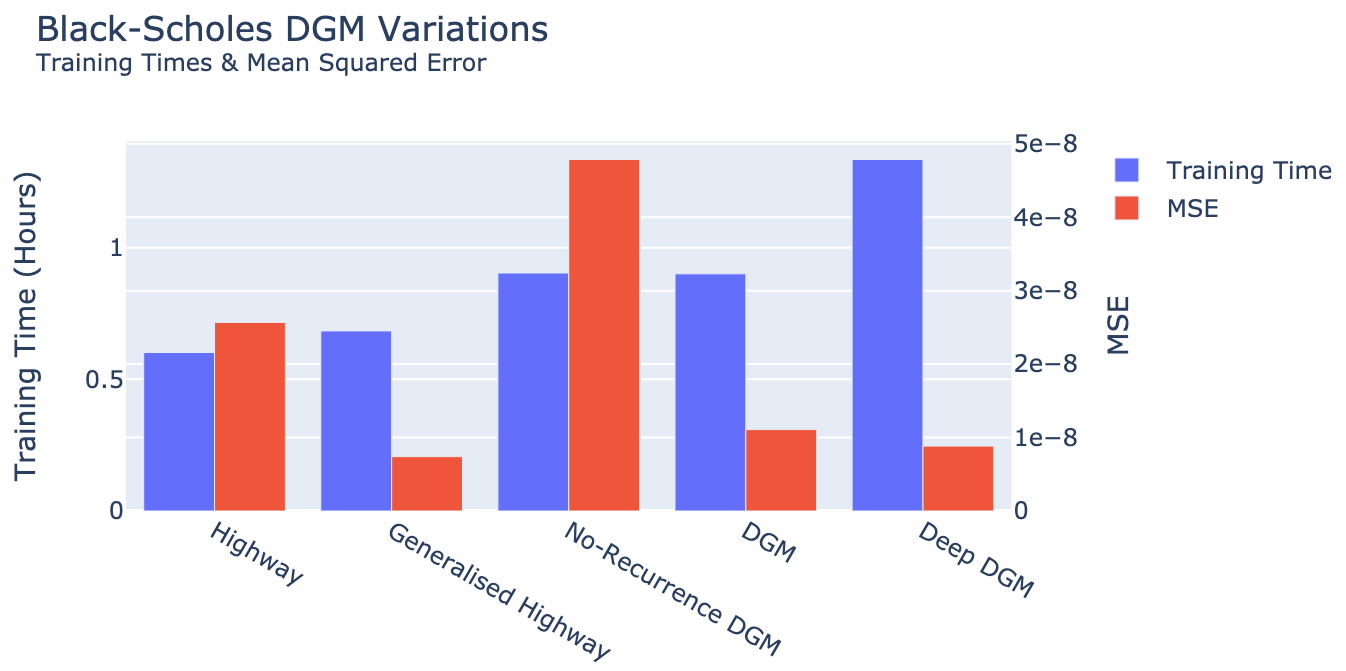}
	\caption{The highway and DGM networks, including DGM variations (deep DGM and no-recurrence DGM) compared in terms of MSE on the test set (red) and training time (blue) for the Black--Scholes pricing problem.}
	\label{fig:bs_dgm_variations_mse_training}
\end{figure}

In order to gain more insights into the impact of the operations inside the DGM network on the performance, we also train the deep DGM network and the no-recurrence DGM network with the same configuration as in Table \ref{table:dgm_configuration}, with the deep DGM layer containing $3$ sublayers. 
In Figure \ref{fig:bs_dgm_variations_mse_training} the MSEs of the DGM variations are compared against the DGM and highway networks. 
As expected from the previous observations, the addition of sublayers inside the DGM layer, resulting in the deep DGM network (\textit{cf.} subsection \ref{subsec:deep_dgm_layer}), does not bring significant performance improvements, while increasing training time by almost $50\%$ compared to the ``plain'' DGM network.
Removing the dependence on the input vector $\mathbf{x}$ in all layers, resulting in the no-recurrence DGM network (\textit{cf.} subsection \ref{subsec:no_recurrence_dgm_network}), more than tripled the network error. 
This suggests that for the Black--Scholes pricing problem, the DGM network admits its performance benefits from the recurrence in the network. 

The operations inside the no-recurrence network are very similar to those in a highway network. 
However, the MSEs on the test set are nowhere close. 
In order to verify that this was not a numerical issue, the no-recurrence network was trained multiple times, yielding similar results. 
The layer operations for the no-recurrence network and the generalized highway network after renaming weight matrices and activation functions are
\begin{align}
	\mathbf{y} = \Big[ 1 - G(\mathbf{x}, W_G) \Big] \odot H\left(R(\mathbf{x}, W_R) \cdot \mathbf{x} \right) + Z(\mathbf{x}, W_Z) \odot \mathbf{x}
\end{align}
and
\begin{align}
	\mathbf{y} = \Big[1 - Z(\mathbf{x}, W_Z)\Big] \odot H(\mathbf{x}, W_H) + Z(\mathbf{x}, W_Z) \odot \mathbf{x},
\end{align}
respectively. 
Hence, the difference in performance likely originates from the additional non-linearity $R$ or $G$.
The fact that the introduction of two additional operations have such a negative influence in accuracy suggests that choosing an optimal network architecture can be a process subject to high sensitivity.

In addition, for the sake of completeness, Table \ref{table:statistics_bs_dgm} lists the Black--Scholes pricing MSE results for the most important networks, alongside the performance of the networks from \cite{jph-report, pricing_options_implied_vol}. 
We notice that the generalized highway networks achieve the lowest errors among the considered architectures.

\begin{table}[h!]
\centering
\begin{tabular}{ c|c|c|r|c|c } 
\hline
Model & Layers & Nodes & Parameters & Training Time (H) & MSE \\ 
\hline
Small MLP & $3$ & $50$ & $5,401$ & $0.55$ & $1.1 \cdot 10^{-7}$ \\
Highway & $3$ & $50$ & $15,601$ & $0.60$ & $2.6 \cdot 10^{-8}$ \\
\rowcolor{table_yellow}
Generalized Highway & $3$ & $50$ & $23,251$ & $0.68$ & $7.4 \cdot 10^{-9}$ \\
No-Recurrence DGM & $3$ & $50$ & $31,059$ & $0.90$ & $4.8 \cdot 10^{-8}$ \\
DGM & $3$ & $50$ & $33,459$ & $0.90$ & $1.1 \cdot 10^{-8}$ \\
Deep DGM & $3$ & $50$ & $49,959$ & $1.33$ & $8.8 \cdot 10^{-9}$ \\
\rowcolor{table_yellow}
Large MLP & $3$ & $500$ & $504,001$ & $2.47$ & $7.0 \cdot 10^{-9}$ \\
MLP \cite{jph-report} & $6$ & $200$ & $202,201$ & - & $1.3 \cdot 10^{-7}$ \\
DGM \cite{jph-report} & $5$ & $100$ & $210,601$ & - & $1.6 \cdot 10^{-7}$ \\
MLP* \cite{pricing_options_implied_vol} & $4$ & $400$ & $644,401$ & - & $8.2 \cdot 10^{-9}$ \\
\hline
\multicolumn{6}{l}{\footnotesize{*Trained for $3000$ epochs, whereas the other networks are trained for $200$.}}
\end{tabular}
\caption{An overview of MLP, highway and DGM network configurations alongside their training time and MSE for the Black--Scholes pricing problem. The best performing networks are highlighted. We also consider the networks from \cite{jph-report} and \cite{pricing_options_implied_vol}.}
\label{table:statistics_bs_dgm}
\end{table}


\subsubsection{Heston model}

\begin{figure}[h!]
	\centering
	\includegraphics[width=0.8\linewidth]{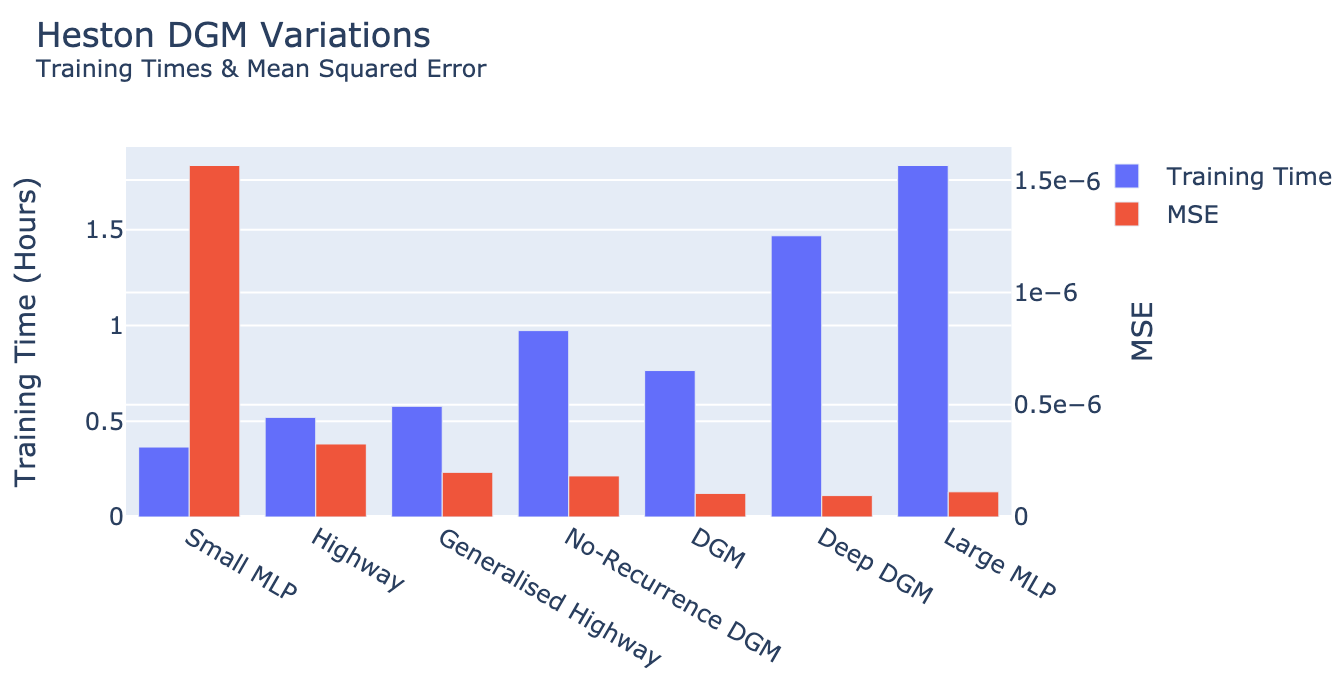}
	\caption{The highway and DGM networks, including DGM variations (deep DGM and no-recurrence DGM) compared in terms of MSE on the test set (red) and training time (blue) for the Heston pricing problem.}
	\label{fig:heston_dgm_variations}
\end{figure}

In Figure \ref{fig:heston_dgm_variations} the performance of the DGM network alongside the smallest and largest $3$-layer MLPs and highway networks in the Heston pricing problem is shown. 
We notice that, unlike  the Black--Scholes problem, the DGM network achieves a lower MSE than both the large MLP and the generalized highway network. 
The reason could be that the DGM has slightly more parameters which can lead to performance improvements; see also the discussion in Section \ref{sec:capacity_normalized}.

Moreover, in the same figure (\cref{fig:heston_dgm_variations}) the results of the deep DGM and the no-recurrence DGM networks are included. 
We notice that neither the no-recurrence network nor the deep DGM network significantly improve performance or training time. 
Interestingly, the no-recurrence DGM also requires slightly longer training time, despite a fewer number of parameters in the network, compared to the DGM network.
We can conclude that the DGM network outperforms both variations. 
In addition, we find that while the recurrence in the DGM network has decreased the MSE, the additional complexity it brings along does not make it preferable for option pricing problems in the tested setup.  

Concluding, we find that the generalized highway network outperforms all other networks in terms of performance compared to total parameters and training time,  similarly to the Black--Scholes problem.


\subsubsection{Implied volatility}

\begin{figure}[h!]
	\centering
	\includegraphics[width=0.8\linewidth]{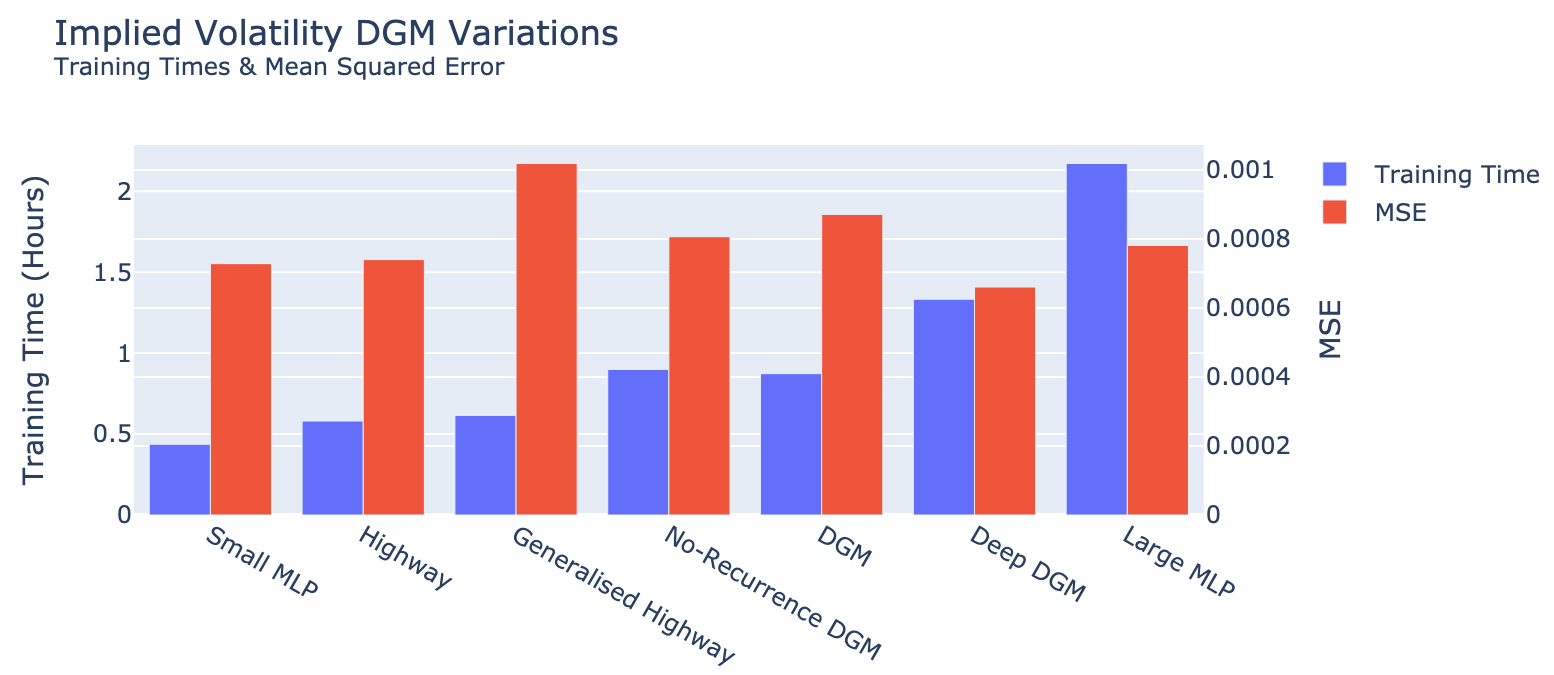}
	\caption{The highway, MLP and DGM network including DGM variations (deep DGM and no-recurrence DGM) compared in terms of MSE on the test set (red) and training time (blue) for the implied volatility problem.}
	\label{fig:iv_deep_dgm_mse_training}
\end{figure}

Finally, we examine the performance of the DGM networks on the implied volatility problem. 
We have seen in the previous subsections that the implied volatility problem has not been susceptible to performance increases with different architectures, which is likely due to the steep gradient problem present in the `default' implied volatility formulation. 
We again find this result when considering the DGM networks. 
Table \ref{table:statistics_iv_dgm} lists the training times and MSE on the implied volatility problem for the networks considered in Figure \ref{fig:iv_deep_dgm_mse_training}. 
We notice that the $3$ layer $150$ node MLP performs best in terms of MSE and training time. 
We find comparable performance from the networks trained in \cite{jph-report, pricing_options_implied_vol}.

\begin{table}[h!]
\centering
\begin{tabular}{ c|c|c|c|c|c } 
\hline
Model & Layers & Nodes & Parameters & Training Time (H) & MSE \\ 
\hline
Small MLP & $3$ & $50$ & $5,401$ & $0.44$ & $7.3 \cdot 10^{-4}$ \\
Highway & $3$ & $50$ & $15,601$ & $0.51$ & $7.4 \cdot 10^{-4}$ \\
Generalized Highway & $3$ & $50$ & $23,251$ & $0.62$ & $1.0 \cdot 10^{-3}$ \\
No-Recurrence DGM & $3$ & $50$ & $31,059$ & $0.90$ & $8.1 \cdot 10^{-4}$ \\
DGM & $3$ & $50$ & $33,459$ & $0.87$ & $8.7 \cdot 10^{-4}$ \\
\rowcolor{table_yellow}
Best MLP & $3$ & $150$ & $46,201$ & $0.60$ & $6.6 \cdot 10^{-4}$ \\
\rowcolor{table_yellow}
Deep DGM & $3$ & $50$ & $49,959$ & $1.33$ & $6.6 \cdot 10^{-4}$ \\
Large MLP & $3$ & $500$ & $504,001$ & $2.17$ & $7.8 \cdot 10^{-4}$ \\
MLP \cite{jph-report} & $6$ & $200$ & $202,201$ & - & $8.1 \cdot 10^{-4}$ \\
\rowcolor{table_yellow}
DGM \cite{jph-report} & $5$ & $100$ & $210,601$ & - & $6.5 \cdot 10^{-4}$ \\
\rowcolor{table_yellow}
MLP* \cite{pricing_options_implied_vol} & $4$ & $400$ & $644,401$ & - & $6.4 \cdot 10^{-4}$ \\
\hline
\multicolumn{6}{l}{\footnotesize{*Trained for $3000$ epochs, whereas the other networks are trained for $200$.}}
\end{tabular}
\caption{An overview of MLP, highway and DGM network configurations along with training time and MSE for the implied volatility problem. The best performing networks are highlighted. We also consider the networks from \cite{jph-report} and \cite{pricing_options_implied_vol}.}
\label{table:statistics_iv_dgm}
\end{table}

\begin{figure}[h!]
	\centering
	\includegraphics[width=0.8\linewidth]{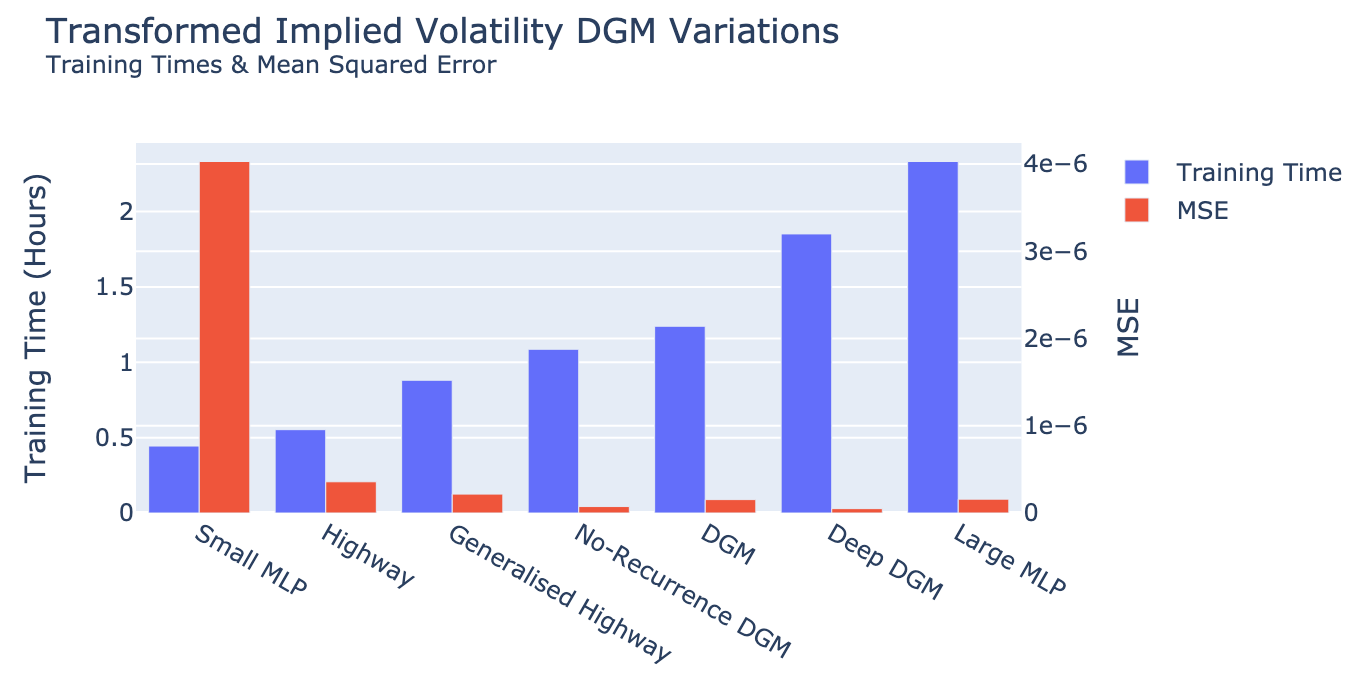}
	\caption{The highway, MLP and DGM networks including DGM variations (deep DGM and no-recurrence DGM) compared in terms of MSE on the test set (red) and training time (blue) for the transformed implied volatility problem.}
	\label{fig:transformed_iv_dgm_variations}
\end{figure}

Since the results we find on the implied volatility dataset is probably a result of convergence problems originating from the formulation, we consider the DGM performance on the transformed implied volatility problem in the remainder of this section. 
We observed in Subsection \ref{subsec:highway_network_analysis} that the highway architectures showed strong signs of decreased MSE with little increase in training time. 
Figure \ref{fig:transformed_iv_dgm_variations} shows the results of the DGM alongside the highway and MLP reference networks. 


We notice that, similar to the Heston pricing problem, the DGM network outperforms the other networks in terms of absolute MSE. 
However, the generalized highway network performs better when the training time is taken into account. 


When we also include the results of the DGM variations, see again Figure \ref{fig:transformed_iv_dgm_variations}, we find that the no-recurrence DGM shows even better performance than the DGM network. 
This network architecture is similar to the generalized highway architecture, but contains one additional non-linear operation. 
In all the previous problems, the additional non-linear operation caused reductions in performance compared to the simpler generalized highway architecture. 
However, for the transformed implied volatility problem, we observe that the no-recurrence architecture shows a $51\%$ reduction in MSE compared to the DGM network, and a $66\%$ reduction compared to the generalized highway. 
Table \ref{table:statistics_transformed_iv_dgm} shows the MSEs, training times and parameters for all the networks of interest.
Taking the training times into account, we can conclude that the no-recurrence DGM network shows the most favourable performance among the architectures considered for the implied volatility problem.

\begin{table}[h!]
\centering
\begin{tabular}{ c|c|c|c|c|c } 
\hline
Model & Layers & Nodes & Parameters & Training Time (H) & MSE \\ 
\hline
Small MLP & $3$ & $50$ & $5,401$ & $0.44$ & $4.0 \cdot 10^{-6}$ \\
Highway & $3$ & $50$ & $15,601$ & $0.55$ & $3.6 \cdot 10^{-7}$ \\
Generalized Highway & $3$ & $50$ & $23,251$ & $0.88$ & $2.2 \cdot 10^{-7}$ \\
\rowcolor{table_yellow}
No-Recurrence DGM & $3$ & $50$ & $31,059$ & $1.09$ & $7.4 \cdot 10^{-8}$ \\
DGM & $3$ & $50$ & $33,459$ & $1.24$ & $1.5 \cdot 10^{-7}$ \\
\rowcolor{table_yellow}
Deep DGM & $3$ & $50$ & $49,959$ & $1.33$ & $5.0 \cdot 10^{-8}$ \\
Large MLP & $3$ & $500$ & $504,001$ & $2.33$ & $1.6 \cdot 10^{-7}$ \\
\rowcolor{table_yellow}
MLP* \cite{pricing_options_implied_vol} & $4$ & $400$ & $644,401$ & - & $1.6 \cdot 10^{-8}$ \\
\hline
\multicolumn{6}{l}{\footnotesize{*Trained for $3000$ epochs, whereas the other networks are trained for $200$.}}
\end{tabular}
\caption{An overview of MLP, highway and DGM network configurations along with training time and MSE for the transformed implied volatility problem. The best performing networks are highlighted. We also consider the network from \cite{pricing_options_implied_vol}.}
\label{table:statistics_transformed_iv_dgm}
\end{table}

To summarise the results of the architectural variations within the DGM family, Table \ref{tab:DGM_variants_final_summary} provides a concise comparison of their structural differences, parameter counts, and empirical performance.

\begin{table}[h!]
\centering
\begin{tabular}{ c|c|c|c|c|c } 
\hline
Model & Layers & Nodes & Parameters & Training Time (H) & MSE \\ 
\hline
No-Recurrence DGM & $3$ & $50$ & $31,059$ & $0.90$ & $4.8 \cdot 10^{-8}$ \\
DGM & $3$ & $50$ & $33,459$ & $0.90$ & $1.1 \cdot 10^{-8}$ \\
Deep DGM & $3$ & $50$ & $49,959$ & $1.33$ & $8.8 \cdot 10^{-9}$ \\
\hline
\end{tabular}
\caption{Comparison of DGM variants in terms of structural features, parameter count, training time, and MSE for the Black-Scholes problem.}\label{tab:DGM_variants_final_summary}
\end{table}


\subsection{Capacity-normalized comparison of architectures}
\label{sec:capacity_normalized}

Although architectural comparisons across networks such as the residual, highway, generalized highway, and DGM networks are informative, the differences in parameter counts can bias conclusions in favour of higher-capacity models. 
Therefore, in order to isolate the effect of the architectural design from the raw model capacity, we conducted a capacity-normalized experiment in which all architectures were adjusted to contain approximately the same number of trainable parameters.

\subsubsection{Experimental setup.} 

We designed new experiments in which all architectures were resized to contain roughly the same number of parameters.  
The following parameter-matched configurations were used:
\begin{itemize}
    \item \textbf{MLP:} 2 layers $\times$ 150 units 
    \item \textbf{Residual:} 3 layers $\times$ 120 units 
    \item \textbf{Highway:} 4 layers $\times$ 50 units 
    \item \textbf{Generalized Highway:} 3 layers $\times$ 50 units 
    \item \textbf{DGM:} 2 layers $\times$ 50 units 
\end{itemize}

All models were trained under identical conditions: Adam optimizer with learning rate $10^{-5}$, batch size of 64, and 200 epochs. 
The same training, validation, and test splits as in the main experiments were used (\textit{i.e.} 80\% / 20\% / 10\%), and each experiment was repeated with three random seeds to report mean and standard deviation for both mean squared error (MSE) and training time.
Tables~\ref{tab:bs_capacity}, \ref{tab:heston_capacity}, and \ref{tab:iv_capacity} summarize the results for the Black--Scholes, Heston, and implied volatility problems, respectively.

\begin{table}[h!]
\centering
\begin{tabular}{lcccccc} 
\hline
Model & Layers & Nodes & Parameters & Training Time (h) & MSE \\ 
      &        &       &            & Mean $\pm$ Std & Mean $\pm$ Std \\
\hline
MLP & 2 & 150 & 23,551 & $1.03 \pm 0.016$ & $(2.76 \pm 0.61)\cdot 10^{-6}$ \\
Residual & 3 & 102 & 22,881 & $1.04 \pm 0.025$ & $(1.00 \pm 0.033)\cdot 10^{-6}$ \\
Highway & 4 & 59 & 22,861 & $1.07 \pm 0.023$ & $(9.97 \pm 5.15)\cdot 10^{-8}$ \\
Generalized Highway & 3 & 59 & 22,421 & $1.08 \pm 0.025$ & $(1.69 \pm 0.27)\cdot 10^{-7}$ \\
DGM & 2 & 50 & 22,459 & $1.01 \pm 0.009$ & $(2.38 \pm 1.74)\cdot 10^{-7}$ \\
\hline
\end{tabular}
\caption{Capacity-normalized comparison for the Black--Scholes problem (all networks have approximately 23k parameters).}
\label{tab:bs_capacity}
\end{table}

\begin{table}[h!]
\centering
\begin{tabular}{lccccc}
\hline
Model & Layers & Nodes & Parameters & Training Time (h) & MSE \\
      &        &       &            & Mean $\pm$ Std & Mean $\pm$ Std \\
\hline
MLP 
& 2 & 150 & 24{,}151 
& $0.92 \pm 0.05$ 
& $(6.2 \pm 5.0)\cdot 10^{-6}$ \\

Residual 
& 3 & 104 & 23{,}713 
& $0.95 \pm 0.02$ 
& $(1.48 \pm 0.05)\cdot 10^{-5}$ \\

Highway 
& 4 & 59 & 22{,}834 
& $1.04 \pm 0.04$ 
& $(2.81 \pm 0.35)\cdot 10^{-5}$ \\

Generic Highway 
& 3 & 59 & 23{,}365 
& $1.06 \pm 0.05$ 
& $(2.31 \pm 0.18)\cdot 10^{-5}$ \\

DGM 
& 2 & 50 & 24{,}467 
& $1.10 \pm 0.02$ 
& $(3.31 \pm 0.16)\cdot 10^{-5}$ \\

\hline
\end{tabular}
\caption{Capacity-normalized comparison for the Heston problem (all networks have approximately 23k parameters).}
\label{tab:heston_capacity}
\end{table}


\begin{table}[h!]
\centering
\begin{tabular}{lcccccc}
\hline
{Model} & {Layers} & {Nodes} & {Parameters} & {Training Time (h)} & {MSE} \\
 & & & & Mean $\pm$ Std & Mean $\pm$ Std \\
\hline
MLP & 2 & 150 & 23,551 & $1.09 \pm 0.01$ & $(1.99 \pm 0.01)\cdot 10^{-3}$ \\
Residual & 3 & 102 & 22,033 & $1.04 \pm 0.07$ & $(3.35 \pm 0.10)\cdot 10^{-3}$ \\
Highway & 4 & 59 & 22,126 & $1.11 \pm 0.01$ & $(2.36 \pm 0.12)\cdot 10^{-3}$ \\
Generalized Highway & 3 & 59 & 22,421 & $1.12 \pm 0.02$ & $(3.32 \pm 0.24)\cdot 10^{-3}$ \\
DGM & 2 & 50 & 22,459 & $2.15 \pm 0.02$ & $(2.04 \pm 0.13)\cdot 10^{-3}$ \\
\hline
\end{tabular}
\caption{Capacity-normalized comparison for the implied volatility problem (all networks have approximately 23k parameters).}
\label{tab:iv_capacity}
\end{table}

\subsubsection{Discussion.}

The capacity-normalized results show, first of all, that the training time is determined by the number of parameters and not the architecture. 
Indeed, as expected, all models in the capacity-normalized experiment need approximately the same time to be trained. 
However, we still observe that architectural differences lead to substantial variation in performance across the three problems. 
In the Black–Scholes setting, highway-type architectures achieve the highest accuracy, indicating that gating mechanisms are well suited for this problem; this is largely consistent with the results of the previous subsections (see \cref{fig:bs_highway_training_set}). 
On the other hand, for the more complex Heston model, the simple MLP yields the lowest error, suggesting that the inductive biases of gated or residual architectures may be overly restrictive under a fixed parameter budget. 
In addition, for the implied-volatility problem, DGM and MLP architectures perform similarly and outperform residual-type models.

Our results show that there is no single best neural network architecture for all problems: each type of network works well for some tasks but not for others.
Importantly, even when all models have the same number of parameters, the differences in performance did not disappear. This means that the results are not caused by some models being bigger or having more capacity. Instead, the differences come from the way each architecture is built.


\subsection{Application to real data and comparison with other methods}

In order to complete the analysis, we evaluated the architectures on real implied volatility data obtained from the OptionMetrics database. 
Each daily implied volatility surface is given on a fixed grid of \emph{maturities}
\[
\{10,\, 30,\, 60,\, 91,\, 122,\, 152,\, 182,\, 273,\, 365\}\ \text{days},
\]
and \emph{put deltas}
\[
\{-90,\,-80,\,-70,\,-60,\,-50,\,-40,\,-30,\,-20,\,-10\}.
\]
Each column of the dataset corresponds to one maturity--delta pair. 
We train the architectures with the same layers and nodes as in Table \ref{table:statistics_iv_dgm}; for the MLP architecture, we consider the best one. 
Table \ref{table:real_iv} reports the results for all five architectures on the real implied volatility data. 
The performance on real implied volatility data is highly consistent with the synthetic experiments shown in Table \ref{table:statistics_iv_dgm}.
In both cases, the MSE falls in the range of $10^{-3}$ to $10^{-4}$. This indicates that the real dataset, although noisy, does not significantly increase the overall approximation difficulty compared to the synthetic implied-volatility surfaces. 
The ranking of architectures also follows similar patterns. The real-market experiment therefore provides an important robustness check: despite data imperfections, neural architectures achieve error levels comparable to the synthetic benchmark.
\begin{table}[h!]
\centering
\begin{tabular}{ c|c|c|c|c|c } 
\hline
Model & Layers & Nodes & Parameters & Training Time (H) & MSE \\ 
\hline
MLP & 3 & 150 & 46,051 & 0.12 & $3.44\cdot 10^{-4}$ \\
Residual & 3 & 50 & 5,501 & 0.12 & $5.73\cdot 10^{-4}$ \\
Highway & 3 & 50 & 10,801 & 0.15 & $3.80\cdot 10^{-4}$ \\
Generalized Highway & 3 & 50 & 16,101 & 0.13 & $3.89\cdot 10^{-4}$ \\
DGM & 3 & 50 & 32,757 & 0.21 & $4.64\cdot 10^{-4}$ \\
\hline
\end{tabular}
\caption{Performance of MLP, highway, and DGM architectures on real implied volatility data from OptionMetrics.}
\label{table:real_iv}
\end{table}

Finally, it is useful to compare the neural-network approaches with classical non-learning methods for computing implied volatility. 
Numerical techniques such as Jäckel’s algorithm, or Chebyshev interpolation can achieve extremely small absolute volatility errors (from about $10^{-6}$ to $10^{-12}$) when applied to well-behaved price inputs; see \citet{glau2017chebyshevmethodimpliedvolatility}. 
These methods are highly precise in controlled settings but do not directly approximate the entire volatility surface across strikes and maturities; instead they apply different schemes depending on the regime of the input (\textit{e.g.} deep in-the-money, \textit{etc}). 
Neural networks, by contrast, aim to learn the full surface in a single model, capturing complex patterns across both dimensions. 
As a result, the neural networks considered in this work reach mean absolute volatility errors around $10^{-2}$, reflecting the intrinsic difficulty of modelling a more irregular and high-dimensional function. 
This comparison highlights that, while classical solvers provide a benchmark for precision, neural networks handle a broader approximation problem.


\section{Conclusion}
\label{sec:conclusion}

The empirical results obtained in this work allow us to draw several practical conclusions regarding the choice of neural network architectures for option pricing and implied volatility computation:
\begin{itemize}
	\item Across the Black--Scholes, Heston, and transformed implied volatility problems, we observe that increasing the number of parameters generally improves accuracy when combined with an appropriate architecture. 
        Larger models typically achieve lower MSEs, although the rate of improvement depends strongly on the underlying problem.
        However, they also require more training time.
    \item Across all architectures, training and validation losses evolve consistently, indicating stable optimization and good generalization without evidence of overfitting. 
        Although validation losses exhibit higher variance, their overall behaviour closely follows the training performance.
	\item The validation loss was often lower during training than the training loss. 
		This indicates that we could improve the accuracy of the networks by training them for more epochs. 
		In \citet{pricing_options_implied_vol}, the results on the Heston pricing problem were more accurate, training for $3000$ cycles compared to $200$ in this work, while using an identical configuration for the MLPs. 
    \item We observe that highway-type network architectures outperformed the other networks for the Black–Scholes problem, when we compared the MSE relative to the training time.
        This was consistent also in the capacity-normalized experiment.
        The same observation is true for the Heston model, but was not consistently observed in the capacity-normalized experiment.   
	\item In the transformed implied volatility problem, the generalized highway network also performed well, but was outperformed by the no-recurrence DGM architecture. 
        Interestingly, this architecture did not perform competitively on the pricing problems, highlighting that architectural benefits are task-dependent.
	\item When training networks on the implied volatility dataset, we found poor performance for all networks, and no improvements when using more complex architectures. 
		When transforming the dataset as explained in Section \ref{subsec:implied_volatility_problem_description}, the prediction accuracy increased significantly, and we found similar results in network architecture performance compared to the Black--Scholes and Heston pricing problems. 
		This observation suggests that the training of networks is highly dependent on the formulation and the dataset. 
		When training neural networks, we should try to ensure that gradients are bounded. 
		A possible way to do this is to transform the input variables so that the gradients with respect to them do not approach zero or infinity.
	\item We found that for the considered problems, the recurrent behaviour of the DGM networks did not yield benefits when compared to the generalized highway networks. 
		Instead, the computational complexity caused by the increase of operations inside the DGM layers made it less attractive in terms of relative performance to off-line required computational power than the generalized highway network. 
    \item The evaluation on real implied volatility data shows that the architectures maintain error magnitudes and relative performance patterns comparable to the synthetic data experiments, indicating that the task-dependent behaviours observed throughout the study persist under real-market conditions.
\end{itemize}   


\appendix

\section{Summary of network architectures}
This appendix presents a comprehensive summary of all neural network configurations employed in our experiments. For each architecture — the MLP, residual, highway, generalized highway, deep DGM, and no-recurrence DGM models — we report:
\begin{itemize}
\item the number of layers and nodes per layer,
\item the activation functions,
\item the total number of trainable parameters,
\item the corresponding references to the relevant tables in the main text. 
\end{itemize}
For all the architectures we used :
\begin{itemize}
    \item loss = MSE,
    \item learning rate = $10^{-5}$,
    \item batch size = 64,
    \item optimizer = Adam,
    \item epochs= 200.
\end{itemize}

\begin{table}[h!]
\centering

\label{tab:feedforward_architectures}
\begin{tabular}{lcccccccc} 
\hline
Architecture & Layers & Nodes & Activation & Initializer & Parameters & References \\ 
\hline
MLP & 2, 3 & 50–500 & ReLU & Glorot  & 2,851-504,001 & \ref{table:mlp_statistics_twelve_models_bs}, \ref{table:mlp_statistics_twelve_models_heston}, \ref{table:mlp_statistics_twelve_models_iv}, \ref{table:mlp_statistics_twelve_models_transformed_iv} \\
Residual & 3 & 50 & $\tanh$ & Glorot & 7,951 & \ref{table:highway_performance_overview}, \ref{table:highway_performance_overview_heston}, \ref{fig:iv_highway_training_set}\\
Highway & 3 & 50 & $\tanh$ & Glorot  & 15,601 & \ref{table:highway_performance_overview}, \ref{table:highway_performance_overview_heston}, \ref{fig:iv_highway_training_set} \\
Generalized Highway & 3 & 50 & $\tanh$ & Glorot & 23,251& \ref{table:highway_performance_overview}, \ref{table:highway_performance_overview_heston}, \ref{fig:iv_highway_training_set} \\
\hline
\end{tabular}
\caption{Overview of feedforward and highway neural network architectures and corresponding hyperparameter configurations.}
\label{tab:feedforward_appendix}
\end{table}
\begin{table}[h!]
\centering

\label{tab:dgm_variants}
\begin{tabular}{lcccccccc} 
\hline
Architecture & Layers & Nodes & Activation & Initializer & Parameters & References\\ 
\hline
DGM & 3 & 50 & $\tanh$ & Glorot Normal & 33,459& \ref{table:statistics_bs_dgm},\ref{table:statistics_iv_dgm},\ref{table:statistics_transformed_iv_dgm}  \\
Deep DGM & 5 & 50 & $\tanh$ & Glorot Normal & 49,959 & \ref{table:statistics_bs_dgm},\ref{table:statistics_iv_dgm},\ref{table:statistics_transformed_iv_dgm}\\
No-Recurrence DGM & 3 & 50 & $\tanh$ & Glorot Normal & 31,059& \ref{table:statistics_bs_dgm},\ref{table:statistics_iv_dgm},\ref{table:statistics_transformed_iv_dgm} \\
\hline
\end{tabular}
\label{tab:dgm_appendix}
\caption{Overview of DGM variants and corresponding hyperparameter configurations.}
\end{table}







\bibliographystyle{abbrvnat}
\bibliography{report} 

\begin{thebibliography}{31}
\providecommand{\natexlab}[1]{#1}
\providecommand{\url}[1]{\texttt{#1}}
\expandafter\ifx\csname urlstyle\endcsname\relax
  \providecommand{\doi}[1]{doi: #1}\else
  \providecommand{\doi}{doi: \begingroup \urlstyle{rm}\Url}\fi

\bibitem[Abadi et~al.(2015)Abadi, Agarwal, Barham, Brevdo, Chen, Citro,
  Corrado, Davis, Dean, Devin, Ghemawat, Goodfellow, Harp, Irving, Isard, Jia,
  Jozefowicz, Kaiser, Kudlur, Levenberg, Man\'{e}, Monga, Moore, Murray, Olah,
  Schuster, Shlens, Steiner, Sutskever, Talwar, Tucker, Vanhoucke, Vasudevan,
  Vi\'{e}gas, Vinyals, Warden, Wattenberg, Wicke, Yu, and
  Zheng]{tensorflow2015-whitepaper}
M.~Abadi, A.~Agarwal, P.~Barham, E.~Brevdo, Z.~Chen, C.~Citro, G.~S. Corrado,
  A.~Davis, J.~Dean, M.~Devin, S.~Ghemawat, I.~Goodfellow, A.~Harp, G.~Irving,
  M.~Isard, Y.~Jia, R.~Jozefowicz, L.~Kaiser, M.~Kudlur, J.~Levenberg,
  D.~Man\'{e}, R.~Monga, S.~Moore, D.~Murray, C.~Olah, M.~Schuster, J.~Shlens,
  B.~Steiner, I.~Sutskever, K.~Talwar, P.~Tucker, V.~Vanhoucke, V.~Vasudevan,
  F.~Vi\'{e}gas, O.~Vinyals, P.~Warden, M.~Wattenberg, M.~Wicke, Y.~Yu, and
  X.~Zheng.
\newblock {TensorFlow}: Large-scale machine learning on heterogeneous systems,
  2015.
\newblock URL \url{https://www.tensorflow.org/}.

\bibitem[Al-Aradi et~al.(2018)Al-Aradi, Correia, Naiff, Jardim, and
  Saporito]{alaradi2018solving}
A.~Al-Aradi, A.~Correia, D.~Naiff, G.~Jardim, and Y.~Saporito.
\newblock Solving nonlinear and high-dimensional partial differential equations
  via deep learning.
\newblock Preprint, arXiv:1811.08782, 2018.

\bibitem[Becker et~al.(2019)Becker, Cheridito, and Jentzen]{MR3960928}
S.~Becker, P.~Cheridito, and A.~Jentzen.
\newblock Deep optimal stopping.
\newblock \emph{Journal of Machine Learning Research}, 20:\penalty0 Paper No.
  74, 25, 2019.

\bibitem[Black and Scholes(1973)]{black_scholes_pde}
F.~Black and M.~Scholes.
\newblock The pricing of options and corporate liabilities.
\newblock \emph{Journal of Political Economy}, 81:\penalty0 637--654, 1973.

\bibitem[Boudabsa and Filipovi\'{c}(2022)]{MR4397928}
L.~Boudabsa and D.~Filipovi\'{c}.
\newblock Machine learning with kernels for portfolio valuation and risk
  management.
\newblock \emph{Finance \& Stochastics}, 26:\penalty0 131--172, 2022.

\bibitem[Buehler et~al.(2019)Buehler, Gonon, Teichmann, and Wood]{MR3977742}
H.~Buehler, L.~Gonon, J.~Teichmann, and B.~Wood.
\newblock Deep hedging.
\newblock \emph{Quantitative Finance}, 19:\penalty0 1271--1291, 2019.

\bibitem[Capponi and Lehalle(2023)]{capponi_lehalle_2023}
A.~Capponi and C.~Lehalle, editors.
\newblock \emph{Machine Learning and Data Sciences for Financial Markets: A
  Guide to Contemporary Practices}.
\newblock Cambridge University Press, 2023.

\bibitem[Cuchiero et~al.(2020)Cuchiero, Khosrawi, and Teichmann]{risks8040101}
C.~Cuchiero, W.~Khosrawi, and J.~Teichmann.
\newblock A generative adversarial network approach to calibration of local
  stochastic volatility models.
\newblock \emph{Risks}, 8\penalty0 (4), 2020.

\bibitem[{D}elft {H}igh {P}erformance {C}omputing~{C}entre
  ({DHPC})(2022)]{DHPC2022}
{D}elft {H}igh {P}erformance {C}omputing~{C}entre ({DHPC}).
\newblock {D}elft{B}lue {S}upercomputer ({P}hase 1).
\newblock \url{https://www.tudelft.nl/dhpc/ark:/44463/DelftBluePhase1}, 2022.

\bibitem[Eberlein et~al.(2010)Eberlein, Glau, and
  Papapantoleon]{fourier_transform_valuation}
E.~Eberlein, K.~Glau, and A.~Papapantoleon.
\newblock Analysis of {F}ourier transform valuation formulas and applications.
\newblock \emph{Applied Mathematical Finance}, 17:\penalty0 211--240, 2010.

\bibitem[Eckstein and Kupper(2021)]{MR4239795}
S.~Eckstein and M.~Kupper.
\newblock Computation of optimal transport and related hedging problems via
  penalization and neural networks.
\newblock \emph{Applied Mathematics \& Optimization}, 83\penalty0 (2):\penalty0
  639--667, 2021.

\bibitem[Fang and Oosterlee(2008)]{Fang_Oosterlee_2008}
F.~Fang and C.~W. Oosterlee.
\newblock A novel pricing method for {E}uropean options based on
  {F}ourier-cosine series expansions.
\newblock \emph{SIAM Journal on Scientific Computing}, 31:\penalty0 826--848,
  2008.

\bibitem[Filipovi{\'c}(2009)]{Filipovic09}
D.~Filipovi{\'c}.
\newblock \emph{{Term-{S}tructure {M}odels: {A} {G}raduate {C}ourse}}.
\newblock Springer, 2009.

\bibitem[Gatheral(2006)]{Gatheral_2006}
J.~Gatheral.
\newblock \emph{The Volatility Surface: A Practitioner's Guide}.
\newblock Wiley, 2006.

\bibitem[Glau et~al.(2019)Glau, Herold, Madan, and
  Pötz]{glau2017chebyshevmethodimpliedvolatility}
K.~Glau, P.~Herold, D.~B. Madan, and C.~Pötz.
\newblock The {C}hebyshev method for the implied volatility.
\newblock \emph{Journal of Computational Finance}, 23\penalty0 (3):\penalty0
  1--31, 2019.

\bibitem[Glorot and Bengio(2010)]{glorot_initializer}
X.~Glorot and Y.~Bengio.
\newblock Understanding the difficulty of training deep feedforward neural
  networks.
\newblock In Y.~W. Teh and M.~Titterington, editors, \emph{Proceedings of the
  Thirteenth International Conference on Artificial Intelligence and
  Statistics}, pages 249--256. PMLR, 2010.

\bibitem[He et~al.(2015)He, Zhang, Ren, and Sun]{he2015delving}
K.~He, X.~Zhang, S.~Ren, and J.~Sun.
\newblock Delving deep into rectifiers: Surpassing human-level performance on
  imagenet classification.
\newblock In \emph{2015 IEEE International Conference on Computer Vision
  (ICCV)}, pages 1026--1034. IEEE Computer Society, 2015.

\bibitem[He et~al.(2016)He, Zhang, Ren, and Sun]{residual_networks}
K.~He, X.~Zhang, S.~Ren, and J.~Sun.
\newblock Deep residual learning for image recognition.
\newblock In \emph{2016 IEEE Conference on Computer Vision and Pattern
  Recognition (CVPR)}, pages 770--778. IEEE Computer Society, 2016.

\bibitem[Heston(1993)]{Heston93}
S.~L. Heston.
\newblock {A closed-form solution for options with stochastic volatility with
  applications to bond and currency options}.
\newblock \emph{Review of Financial Studies}, 6:\penalty0 327--343, 1993.

\bibitem[Horvath et~al.(2021)Horvath, Muguruza, and Tomas]{MR4188878}
B.~Horvath, A.~Muguruza, and M.~Tomas.
\newblock Deep learning volatility: a deep neural network perspective on
  pricing and calibration in (rough) volatility models.
\newblock \emph{Quantitative Finance}, 21:\penalty0 11--27, 2021.

\bibitem[Hur\'{e} et~al.(2021)Hur\'{e}, Pham, Bachouch, and
  Langren\'{e}]{MR4218407}
C.~Hur\'{e}, H.~Pham, A.~Bachouch, and N.~Langren\'{e}.
\newblock Deep neural networks algorithms for stochastic control problems on
  finite horizon: convergence analysis.
\newblock \emph{SIAM Journal on Numerical Analysis}, 59\penalty0 (1):\penalty0
  525--557, 2021.

\bibitem[Liu et~al.(2019)Liu, Oosterlee, and
  Bohte]{pricing_options_implied_vol}
S.~Liu, C.~W. Oosterlee, and S.~M. Bohte.
\newblock Pricing options and computing implied volatilities using neural
  networks.
\newblock \emph{Risks}, 7\penalty0 (1), 2019.

\bibitem[McKay et~al.(1979)McKay, Beckman, and
  Conover]{latin_hypercube_sampling}
M.~D. McKay, R.~J. Beckman, and W.~J. Conover.
\newblock A comparison of three methods for selecting values of input variables
  in the analysis of output from a computer code.
\newblock \emph{Technometrics}, 21\penalty0 (2):\penalty0 239--245, 1979.

\bibitem[Papazoglou-Hennig(2020)]{jph-report}
J.~Papazoglou-Hennig.
\newblock Elementary stochastic calculus, application to financial mathematics,
  and neural network methods for financial models, 2020.
\newblock Internship notes, NTUA-TUM.

\bibitem[Ruf and Wang(2020)]{RW_JCF}
J.~Ruf and W.~Wang.
\newblock Neural networks for option pricing and hedging: a literature review.
\newblock \emph{Journal of Computational Finance}, 24\penalty0 (1):\penalty0
  1--46, 2020.

\bibitem[Sherstinsky(2020)]{Sherstinsky_2020}
A.~Sherstinsky.
\newblock Fundamentals of recurrent neural network ({RNN}) and long short-term
  memory ({LSTM}) network.
\newblock \emph{Physica D: Nonlinear Phenomena}, 404:\penalty0 132306, 28,
  2020.

\bibitem[Sirignano and Spiliopoulos(2018)]{MR3874585}
J.~Sirignano and K.~Spiliopoulos.
\newblock D{GM}: a deep learning algorithm for solving partial differential
  equations.
\newblock \emph{Journal of Computational Physics}, 375:\penalty0 1339--1364,
  2018.

\bibitem[Srivastava et~al.(2015)Srivastava, Greff, and
  Schmidhuber]{srivastava2015highway}
R.~K. Srivastava, K.~Greff, and J.~Schmidhuber.
\newblock Highway networks.
\newblock In \emph{ICML Deep Learning Workshop}, 2015.

\bibitem[Telgarsky(2016)]{Telgarsky2016BenefitsOD}
M.~Telgarsky.
\newblock Benefits of depth in neural networks.
\newblock In V.~Feldman, A.~Rakhlin, and O.~Shamir, editors, \emph{29th Annual
  Conference on Learning Theory}, pages 1517--1539. PMLR, 2016.

\bibitem[Van~Mieghem(2022)]{van_Mieghem_2021}
L.~Van~Mieghem.
\newblock Option pricing techniques using neural networks.
\newblock Master's thesis, TU Delft, 2022.

\bibitem[Zhang et~al.(2022)Zhang, Zohren, and Roberts]{ZZR}
Z.~Zhang, S.~Zohren, and S.~Roberts.
\newblock Deep learning for portfolio optimization.
\newblock \emph{The Journal of Financial Data Science}, 2:\penalty0 8--20,
  2022.

\end{thebibliography}


\end{document}